\documentclass[sigconf, nonacm]{acmart}
\AtBeginDocument{%
  }

\setcopyright{none} 
\copyrightyear{2025}
\acmYear{2025}
\acmDOI{XXXXXXX.XXXXXXX}

\acmConference[MICRO 2025]{The 58th IEEE/ACM International Symposium on Microarchitecture}{October 18--22, 2025}{Seoul, Korea}

\acmISBN{978-X-XXXX-XXXX-X/XX/XX}

\usepackage{float}

\usepackage{stfloats}

\newcommand\ket[1]{\left|#1\right\rangle}

\usepackage{amsmath,mathtools, amsthm}

\newcommand\todoo[1]{\textcolor{black}{#1}}

\newcommand\revise[1]{\textcolor{black}{#1}}

\usepackage[normalem]{ulem}

\usepackage{xcolor}
\usepackage{color}
\definecolor{codegreen}{rgb}{0,0.6,0}
\definecolor{codegray}{rgb}{0.5,0.5,0.5}
\definecolor{codepurple}{rgb}{0.58,0,0.82}
\definecolor{backcolour}{rgb}{0.95,0.95,0.92}
\definecolor{textblue}{rgb}{.2,.2,.7}
\definecolor{textred}{rgb}{0.54,0,0}
\definecolor{textgreen}{rgb}{0,0.43,0}
\definecolor{codered}{rgb}{201,72,12}
\usepackage[T1]{fontenc}
\usepackage[scaled=0.85]{beramono} 
\usepackage{listings}
\usepackage{multirow}
\usepackage{multicol}
\usepackage{url}
\usepackage[ruled,vlined,linesnumbered]{algorithm2e}
\usepackage{multirow}
\usepackage{tikz}
\usepackage{xcolor}
\usepackage{ctable} 

\begin{document}

\title{OneAdapt: Adaptive Compilation for Resource-Constrained Photonic One-Way Quantum Computing}
\author{Hezi Zhang}
\email{hezi@ucsd.edu}
\affiliation{
 \institution{University of California, San Diego}
 \city{La Jolla, California}
 \country{USA}
}

\author{Jixuan Ruan}
\email{j3ruan@ucsd.edu}
\affiliation{
 \institution{University of California, San Diego}
 \city{La Jolla, California}
 \country{USA}
}

\author{Dean Tullsen}
\email{tullsen@ucsd.edu}
\affiliation{
 \institution{University of California, San Diego}
 \city{La Jolla, California}
 \country{USA}
}

\author{Yufei Ding}
\email{yufeiding@ucsd.edu}
\affiliation{
 \institution{University of California, San Diego}
 \city{La Jolla, California}
 \country{USA}
}

\author{Ang Li}
\email{ang.li@pnnl.gov}
\affiliation{
 \institution{Pacific Northwest National Laboratory}
 \city{Richland, Washington}
 \country{USA}
}

\author{Travis S. Humble}
\email{humblets@ornl.gov}
\affiliation{
 \institution{Oak Ridge National Laboratory}
 \city{Oak Ridge, Tennessee}
 \country{USA}
}





\begin{abstract}
Measurement-based quantum computing (MBQC), a.k.a. one-way quantum computing (1WQC), is a universal quantum computing model, which is particularly well-suited for photonic platforms. In this model, computation is driven by measurements on an entangled state, which serves as an intermediate representation (IR) between program and hardware. 
However, compilers on previous IRs lacks the adaptability to the resource constraint in photonic quantum computers.
In this work, we propose a novel IR with new optimization passes. Based on this, it realizes a resource-adaptive compiler that minimizes the required hardware size and execution time while restricting the requirement for fusion devices within an adaptive limit.
Moreover, our optimization can be
integrated with Quantum Error Correction (QEC) to improve the efficiency of photonic fault-tolerant quantum computing (FTQC). 


\end{abstract}


\keywords{Quantum Computing, Photonics, MBQC, 1WQC, Compilation}

\maketitle

\section{Introduction}
\label{sect:introduction}

\begin{figure*}[tp]
    \centering
    \includegraphics[width=\textwidth]{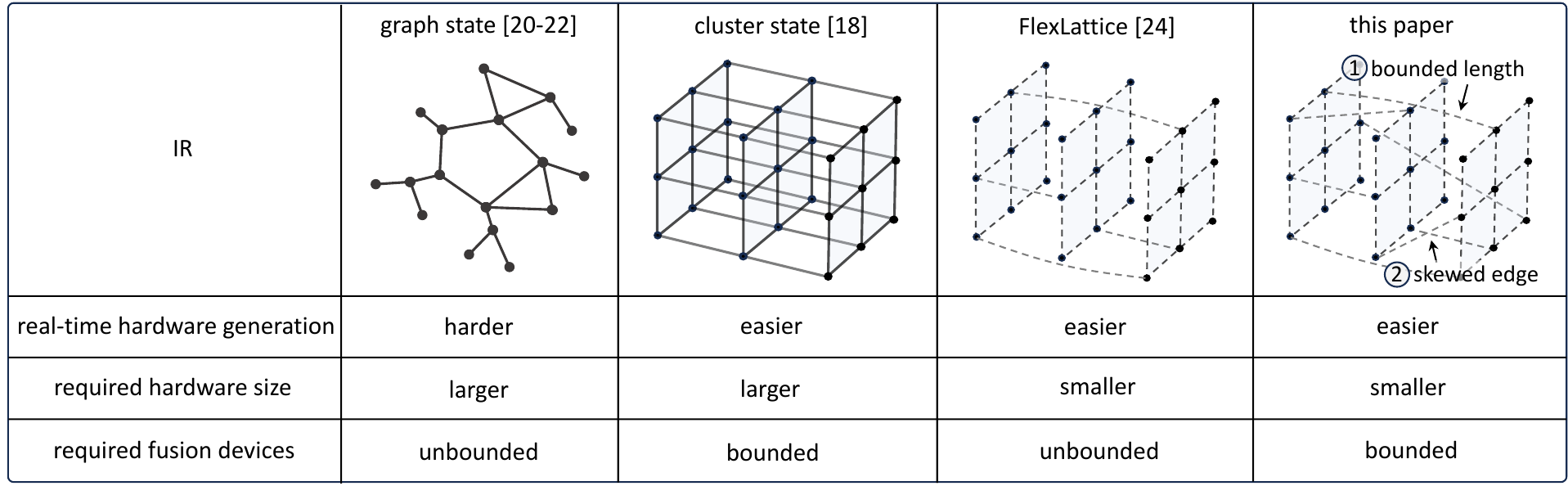}\\
    \caption{Different IRs for photonic one-way quantum computing.}
    \label{fig:IR_table}
\end{figure*}

Photonic systems are a promising platform for universal quantum computing due to the unique advantages of photonic qubits~\cite{OBrien2009,Bogdanov:17}, including their great scalability, long coherence time and easy integration with quantum networks. In addition to experimental demonstrations of quantum supremacy on photonic systems~\cite{zhong2020quantum,PhysRevLett.127.180502,Madsen2022}, significant manufacturing progress has been demonstrated recently on integrated silicon photonics \cite{manufacturable}, achieving a 99.98\% state preparation and measurement fidelity and a 99.22\% fusion fidelity (not to be confused with fusion success rate of $<78\%$~\cite{0.78}). 

In contrast to other quantum computing platforms such as superconducting \cite{SC_1,SC_2,SC_3,SC_4}, trapped ion \cite{TrappedIon_1,TrappedIon_2, TrappedIon_3} and neutral atoms \cite{NA_1,NA_2,NA_3}, photonic platforms are well suited for a different computing model called measurement-based quantum computing (MBQC) or one-way quantum computing (1WQC) \cite{mbqc2009}. In this model, computation is driven by single-qubit measurements on an entangled state. Quantum programs need to be transformed into a measurement pattern on this entangled state, while physical hardware is responsible for generating this entangled state by merging small resource states through \emph{fusions} \cite{fbqc}. As a result, this entangled state can serve as an intermediate representation (IR) between a high-level program and low-level hardware.

Early work in MBQC has transformed programs into ideal IRs without considering hardware capability or cost.
Some approaches assume the availability of an entangled state with arbitrary graph structures, creating an extremely flexible IR known as a graph state \cite{graph_state,determinism,translation}, as shown in Fig.~\ref{fig:IR_table}. Others assume an entangled state with an infinitely large, fixed 3D lattice structure (or 2D \cite{Pitt_MBQC}), resulting in a rigid IR called a cluster state \cite{mbqc2009}, as shown in Fig.~\ref{fig:IR_table}. 
Recent work introduces an approach to generate a FlexLattice IR between these two extremes within a practical hardware model \cite{OnePerc}, structured with two finitely large spatial dimensions and one infinitely long temporal dimension. It maintains a lattice structure within each 2D layer, but allows adjustable edges (dashed lines in Fig.~\ref{fig:IR_table}) and connections between nodes on the same locations between any 2D layers (referred to as temporal edges).

\begin{figure}[h]
    \centering
    \includegraphics[width=\linewidth]{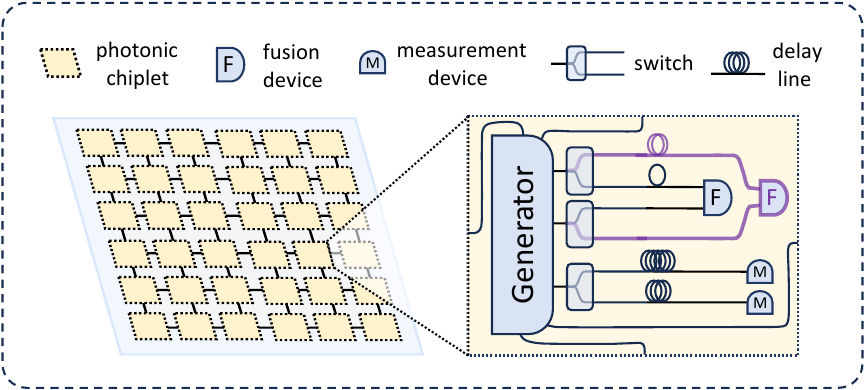}\\
    \caption{Physical hardware of photonic platform.}
    \label{fig:intro}
\end{figure}

These IRs fail to simultaneously support scalable hardware generation and efficient program transformation---two essential requirements for bridging programs and hardware.
For example, due to the probabilistic nature of fusions \cite{0.78,0.95with30}, generating a graph state IR with arbitrary structures at runtime requires endless fusion retrials (Fig.~\ref{fig:IR_table}). While the cluster state IR and the FlexLattice IR avoid this by reducing runtime generation to their rigid lattice structures on each 2D layer, they each come with their own limitations.
For the cluster state IR, program transformation demands a relatively large fixed 2D size, which requires larger hardware size (more chiplets in Fig.~\ref{fig:intro}) and incurs higher runtime overhead (Fig.~\ref{fig:IR_table}).
In contrast, the FlexLattice IR allows program transformation onto a smaller 2D size through its flexible temporal edges, but this flexibility may lead to an unbounded requirement on the number of fusion devices as the program scales (Fig.~\ref{fig:IR_table}).

With this insight, we identify two key opportunities to improve the FlexLattice IR by analyzing the hardware implications of temporal edges and their impact on program transformation. \textbf{First}, we constrain the lengths of temporal edges to avoid unbounded hardware demand. Specifically, supporting edges of arbitrary temporal length from 1 to $n$ requires $O(n)$ fusion devices, each connected to a delay line with a different length \cite{interleaving}. This is because 
temporal edges are realized in hardware by adjusting photon arrival times using delay lines of varying lengths, allowing photons generated at different times to be fused together. 
For example, in Fig.~\ref{fig:intro}, the two purple fibers represent a no-delay and a two-layer delay: photons generated two layers apart can be routed through these fibers to arrive simultaneously at the fusion device.
\textbf{Second}, we introduce skewed temporal edges, which are allowed to connect nodes on different 2D layers within a small 2D distance. While the FlexLattice IR only permits temporal edges between nodes at the same 2D location, we observe that its hardware implementation can accommodate skewed edges with minor modifications. When the skew is bounded to a small 2D range, this flexibility introduces negligible hardware overhead, yet opens up significant opportunities for program optimization.

In this paper, we investigate program transformation with this new IR, proposing a resource-adaptive compiler that minimizes (i) the 1D depth and (ii) the 2D size of the transformed IR program while restricting (iii) the lengths of temporal edges to an adaptive limit.
This is highly non-trivial as it needs to balance between optimization goals that appear conflicted. 
To this end, we propose two novel optimization passes for this new IR. 
\textbf{First}, we employ a dynamic node refresh to restrict the length limit of temporal edges. In contrast to the periodic refresh in OnePerc \cite{OnePerc}, it realizes a more fine-grained refresh of nodes based on their storage time in delay lines, ensuring a refresh before the temporal edge length of each node approaches the limit. With a dynamic adjustment of the percentage of refreshed nodes on each 2D layer, it avoids a significant increase in 1D depth by striking a balance between mapping of new nodes and refresh of old nodes.
\textbf{Second}, we incorporate a 2D-bounded temporal routing to exploit the optimization space enlarged by the skewed edges. 
Specifically, temporal edges between different 2D layers are skewed in suitable directions to bring nodes closer on subsequent layers, with its simplest usecase resembling a native SWAP gate in the circuit model. This reduces the 2D size as well as the 1D depth of IR without necessarily increasing the lengths of temporal edges. 

Our contributions in this paper can be summarized as below:

\begin{itemize}
    \item We propose a new IR by constraining the lengths of temporal edges and allowing skewed edges between 2D layers, which ensures bounded hardware overhead while opening up broader optimization space for program compilation. 
    \item With this new IR, we propose a resource-adaptive compiler with two optimization passes: a dynamic node refresh to ensure the restriction of temporal edge lengths and a 2D-bounded temporal routing to enable more efficient routing with the skewed edges. 
    

    \item Our evaluation shows that, compared to FlexLattice IR compilers, our compiler reduces the 1D depth by \todoo{$3.68\times$} while constraining temporal edge lengths within an adaptive bound (as low as 1). Compared to cluster state IR compilers, it can achieve a \todoo{$3.56\times$} 1D depth reduction or significantly reduce the 2D size (e.g., from $8\times8$ to $3\times3$ for 64-qubit programs). 



    \item We extend our framework from the NISQ setting to FTQC using surface code, demonstrating its ability to integrate with QEC, 
    reducing the 1D depth by a factor of \todoo{2.87$\times$.}

\end{itemize}


\section{Background and Related Work}\label{sect:background}
\subsection{MBQC Basics}
\begin{figure}[h!]
    \centering
    \includegraphics[width=\linewidth]{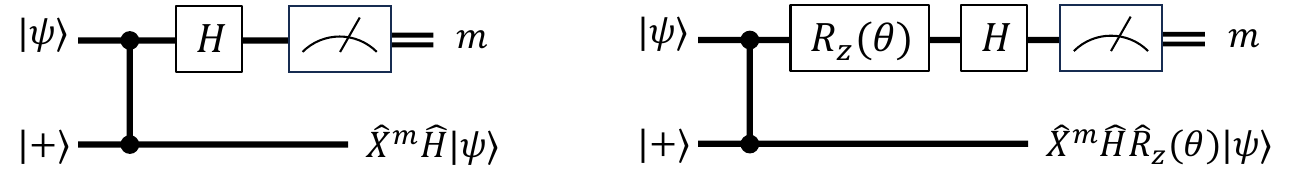}\\
    \hspace{0pt}(a)\hspace{120pt}(b) \hspace{10pt}
    \caption{1-qubit teleportation (a) and its generalization (b).}
    \label{fig:1-qubit_teleportation}
\end{figure}

\textbf{One-qubit Teleporation}
The foundation of MBQC is the well-known concept of 1-qubit teleportation. As shown in Fig.~\ref{fig:1-qubit_teleportation}(a), if two qubits are initialized to $CZ(\ket{\psi}\otimes\ket{+})$, when the first qubit $\ket{\psi}$ is measured in the $X$-basis (implemented by applying a Hadamard gate $H$ before a $Z$-basis measuremet), the state of the second qubit would become $X^m H\ket{\psi}$, where $X$ is a Pauli correction induced by the measurement outcome $m$ of the first qubit. For a measurement in a general basis on the $X$-$Y$ plane of the bloch sphere (implemented by applying a $Z$-rotation $R_z(\theta)$ before the $X$-basis measurement), the state of the second qubit would become $X^m HR_z(\theta)\ket{\psi}$. In these examples, the resulting state of the second qubit is determined by the measurement basis of the first qubit.

\noindent\textbf{Graph State}
Computation in the MBQC paradigm is driven by single-qubit measurements on an entangled state, called \emph{graph state}.
Formally, it is an entangled state among qubits located on a graph $G(V,E)$, defined as the eigenstates of operator \[s=X_i \bigotimes_{j\in n_i} Z_j, \quad\forall i\in V\] 
where $n_i$ are the neighboring qubits of $i\in V$ on graph $G$. A graph state can be created by preparing all qubits in $\ket{+}$, with some input qubits being the input states, and then applying a $CZ$ gate between each pair of qubits connected by a graph edge.
Computation on a graph state can be implemented by a pattern of \emph{equatorial measurements} $E(\alpha)$, i.e., measurements on the $X$-$Y$ plane of Bloch sphere at an angle $\alpha$.
The measurement basis of each qubit is predetermined by the quantum program, together referred to as a \emph{measurement pattern}. But they are subject to a real-time adjustment according to the measurement outcomes of prior qubits, with the angles adjusted from $\alpha$ to $(-1)^s\alpha+t\pi$ where $s,t\in\{0,1\}$. This feed-forward compensates the non-determinism of measurement outcomes and eliminates the Pauli corrections by propagating them to the end of the computation.

MBQC has the same computation power with the circuit model since they are both universal computing models. 
The required structure $G$ of the graph state of a quantum program can be obtained by a straightforward translation~\cite{translation} from a circuit.
 This process can be described in ZX-calculus \cite{zx_calculus} and optimized by available tools such as PyZX~\cite{kissinger2019pyzx}.


\subsection{Photonic Platform} 

\begin{figure*}[tp!]
        \centering
        \includegraphics[width=\linewidth]{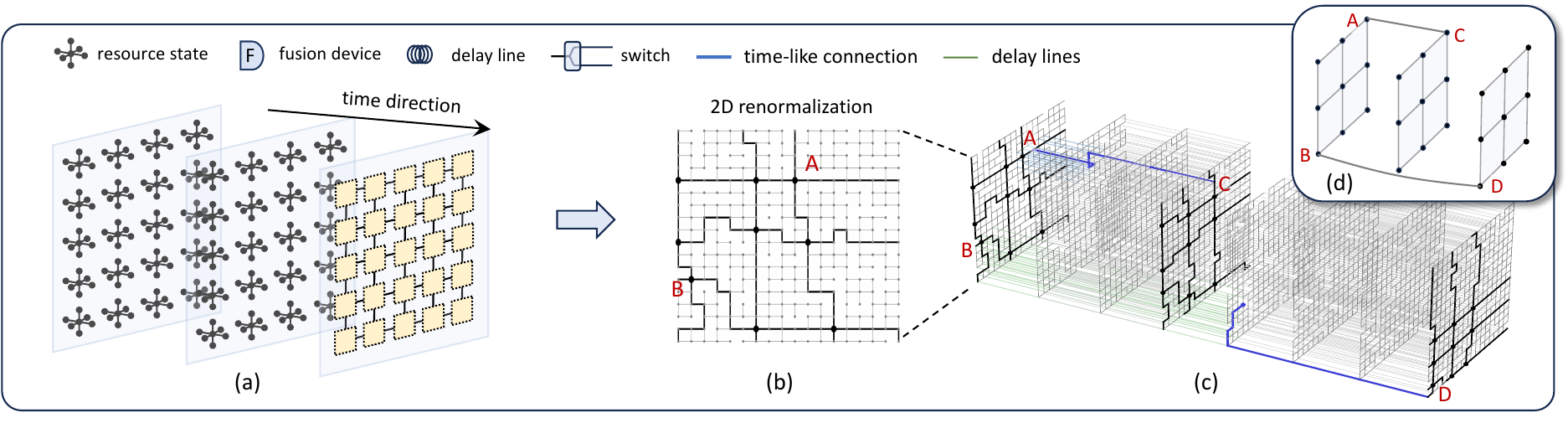}
        
        \caption{Physical hardware of photonic platform}
        \label{fig:fbqc_hardware}
\end{figure*}

\textbf{Fusion-based Architecture}
On photonic platforms, a large graph state can be generated by merging small resource graph states through (type II \cite{linear_optics}) fusions \cite{fbqc},  which are concurrent measurements on two photonic qubits from different resource states in two entangled bases, such as $X\otimes Z$ and $Z\otimes X$.

In the architecture proposed recently by PsiQuantum \cite{fbqc}, a photonic processor consists of an array of interconnected chiplets as shown in Fig.~\ref{fig:fbqc_hardware}(a). Each chiplet can be equipped with a resource state generator (RSG), delay lines, fusion and measurement devices, as shown in Fig.~\ref{fig:fbqc_hardware}(a). Each RSG generates a copy of resource state in every cycle, which will be directed to different fusion or measurement devices by switching between different delay lines. 
Due to connectivity limitations, resource states generated at the same cycle are restricted to fuse with their neighbors, while those generated at different cycles are restricted to fuse with others generated by the same RSG. 

With this architecture, resource states generated in the same RSG cycle form a 2D resource state layer (RSL), with resource state fusions leading to a 3D structure in space-time when the time dimension is incorporated, as shown in Fig.~\ref{fig:fbqc_hardware}(c). Here the resource states are assumed to be 7-qubit star-shaped, as demonstrated earlier in Fig.~\ref{fig:fbqc_hardware}(a). When the available resource states are smaller, this can also be achieved by merging more RSLs \cite{OnePerc}.

\textbf{Experimental Progress} So far, most key hardware components required by this architecture have been demonstrated experimentally, including resource state generation of up to 8 qubits \cite{fusion_nature_2024, ferreira2024deterministic, coste2023high, cogan2023deterministic, thomas2022efficient, yang2022sequential,besse2020realizing,istrati2020sequential,schwartz2016deterministic}, heralded (type II~\cite{linear_optics}) fusions with 99.22\% fidelity \cite{fusion_nature_2024,manufacturable}, single-qubit measurements with 99.98\% fidelity \cite{manufacturable}, 
chip-to-chip qubit interconnect with 99.72\% fidelity \cite{manufacturable} and potential delay lines with a low photon loss rate of 0.2 dB/km \cite{fiber}. Moreover, the feasibility of MBQC has been demonstrated with various small-scale quantum algorithms on photonic qubits \cite{mbqc-grover,mbqc-dj,mbqc-simon}.

\subsection{MBQC on Photonic Platform}
\textbf{MBQC Compilation with IR} Compilation of MBQC can be formulated as a problem of mapping program graph states onto a hardware-friendly graph state with a structure that is easy to generate, which serves as an intermediate representation (IR) that bridges program and hardware. Early work in MBQC \cite{mbqc2009} proposes a cluster state IR with a lattice structure, demonstrating universality of MBQC by translating a universal gate set in the circuit model to corresponding measurement patterns on cluster states. Recently, a compiler FMCC \cite{Pitt_MBQC} is proposed to optimize this translation, with the cluster state restricted to a 2D scenario. Another work OnePerc \cite{OnePerc} proposes a more flexible FlexLattice IR, with a 3D lattice-like structure that allows edges along the third dimension to connect nodes at the same 2D positions of any 2D layers, as depicted in Fig.~\ref{fig:IR_table} and Fig.~\ref{fig:fbqc_hardware}(d). Moreover, each edge in this IR can be enabled or disabled on demand. For compilation, OnePerc adopts a slightly modified mapping algorithm from \cite{OneQ}.


\textbf{IR Generation} 
To generate the FlexLattice IR, resource states on each RSL are fused with their neighbors while different RSLs are fused according to the IR structure by switching between different delay lines, with the formed entanglement depicted by the grey lines in Fig.~\ref{fig:fbqc_hardware}(b)(c) and delay lines depicted by the green lines in Fig.~\ref{fig:fbqc_hardware}(c). To handle the probabilistic failures of fusions \cite{0.78,0.95with30}, which are indicated by the missing edges in Fig.~\ref{fig:fbqc_hardware}(c), a percolation-based \cite{square_lattice_perc,perc_threshold} algorithm can be adopted to ensure that the structure can be generated with a high probability through those successful fusions \cite{OnePerc}. Specifically:

(1) On each RSL, a path searching is conducted among successful fusions to identify a connected lattice of a certain size, which is referred to as \emph{renormalization} \cite{2D_renorm,3D_renorm}, as illustrated in Fig.~\ref{fig:fbqc_hardware}(b) when the lattice size is $3\times 3$. A successfully renormalized RSL can serve as a \emph{logical layer}, corresponding to a 2D layer in the IR.

(2) To generate an IR structure as shown in Fig.~\ref{fig:fbqc_hardware}(d), two time-like edges $AC$ and $BD$ need to be generated. 
This is achieved by finding connected paths between the renormalized nodes $A$ and $C$, and between $B$ and $D$, as depicted by bold blue paths in Fig.~\ref{fig:fbqc_hardware}. Note that since $BD$ connects two non-adjacent 2D layers in the IR, the resource states around $B$ are delayed to fuse with the RSL after the second logical layer, unlike other resource states that are fused with their next RSL.

\textbf{PL Ratio} Due to fusion failures, the formation of each logical layer costs multiple RSLs. This is because both 2D renormalization and time-like path searching can fail. The ratio between the number of physical RSLs and logical layers is thus defined as a \emph{PL ratio}. With a practical fusion success probability of 75\% \cite{0.78,0.95with30}, simulation in \cite{OnePerc} reveals this ratio as $\sim 3.1$ on average, which can vary between different logical layers. In real compilation, this ratio can be fixed to avoid real-time feed-forward overhead. To ensure a high success probability of IR generation, a reasonable choice would be slightly larger than the average ratio, such as 4, meaning that a logical layer is allocated every 4 RSLs.

\subsection{FTQC on Photonic Platform}
To realize photonic FTQC, multiple physical qubits need to be encoded into a logical qubit through quantum error correction (QEC) codes \cite{nielsen2001quantum}. 
Existing work has shown that any QEC code in the CSS category can be implemented on MBQC in a foliated manner \cite{bolt2016foliated}. In this approach, each RSL mimics a time slice of error correction, with the fusion patterns on the RSL determined by the code structure. For surface code, one of the most promising QEC codes \cite{surface_code}, these RSLs are fused into a Raussendorf cluster state \cite{raussendorf2006fault}, with the structure on each of its layers resembling the structure of surface code. This correspondence between an MBQC layer and a time slice of QEC enables implementation of universal logical gates in the MBQC paradigm \cite{Raussendorf_lattice_surgery,FBQC_logical}.

\begin{figure}[h!]
    \centering
    \includegraphics[width=\linewidth]{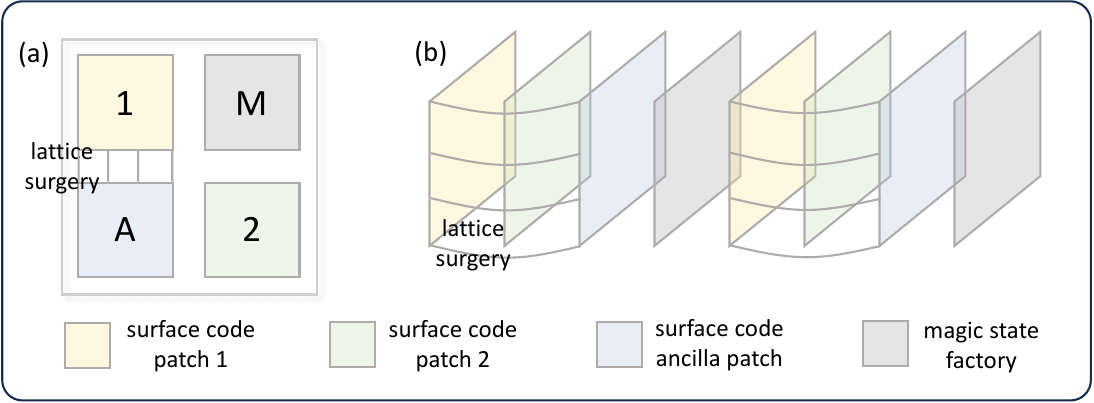}
    \caption{Periodic Refresh for QEC}
    \label{fig:QEC_dynamic}
\end{figure}

Fusion-based architectures require a large QEC code distance due to the need to tolerate the high fusion failure rate and photon loss rate \cite{FBQC_logical}, leading to a requirement for large RSLs. This resource overhead can be mitigated by a space-time trading that allocates different logical qubits onto different RSLs. Fig~\ref{fig:QEC_dynamic}(a) illustrates a simple example where four surface code patches are required, including an ancilla patch and a magic state factory to facilitate logical gates with lattice surgery \cite{lattice_surgery,game,high_perform_large_surface} and magic state distillation \cite{low_overhead, edge_disjoint}. As depicted in Fig.~\ref{fig:QEC_dynamic}(b), these patches can be allocated periodically in an interleaved manner, with the period being four layers here. As a lattice surgery \cite{lattice_surgery,game,high_perform_large_surface} between two patches only involves operations on the patch boundaries (Fig.~\ref{fig:QEC_dynamic}(a)), this can be realized through the boundaries of different layers with the help of delay lines (Fig.~\ref{fig:QEC_dynamic}(b)). 






\begin{figure*}[tp]
    \centering
    \includegraphics[width=\textwidth]{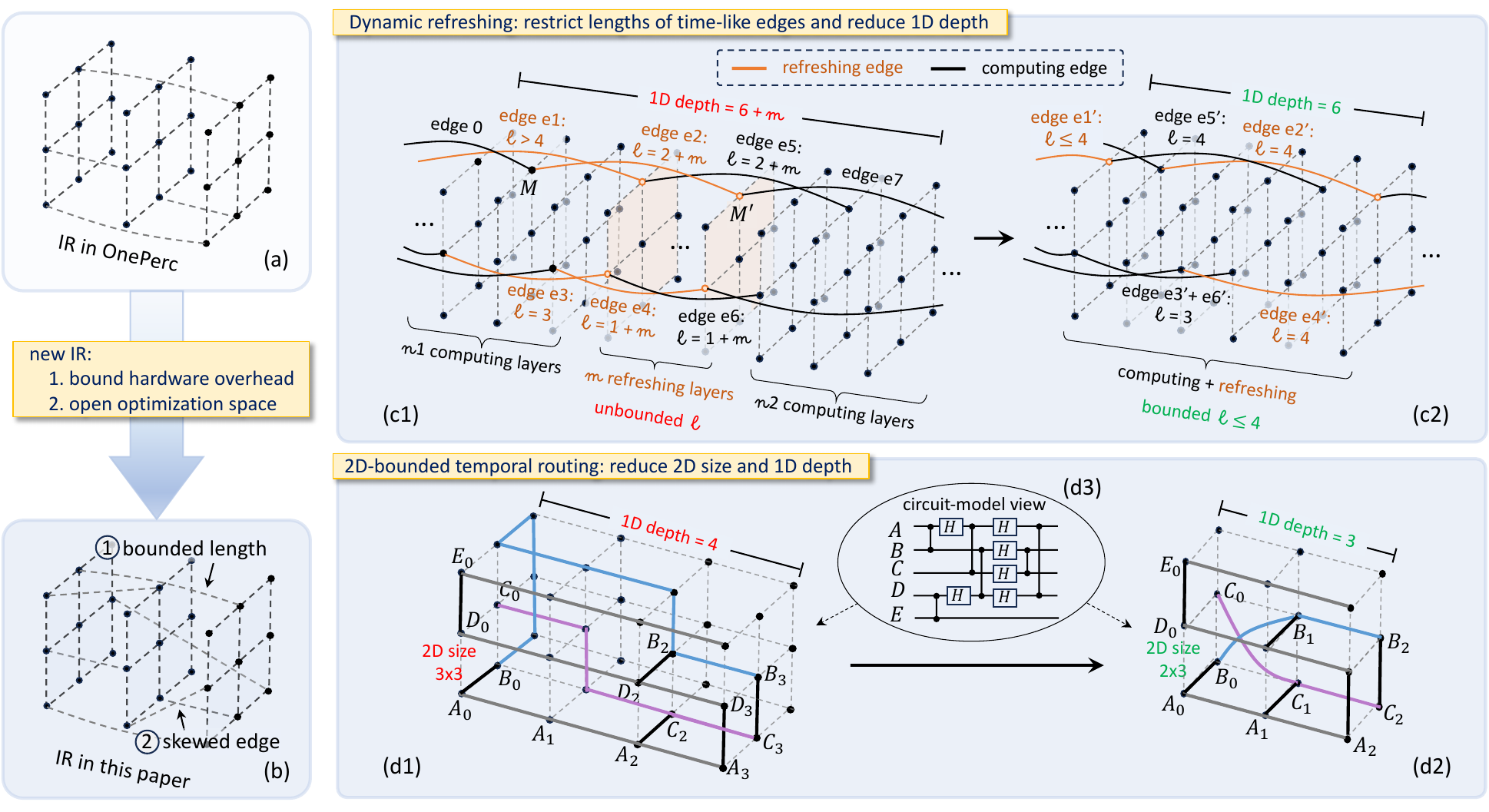}
    \caption{IR in OnePerc (a) and  IR in this paper (b). Dynamic refreshing that prevents temporal edges from exceeding a length limit (c) and 2D-bounded temporal routing that exploits skewed edges in the IR (d).}
    \label{fig:overview_example}
\end{figure*}

\section{Design Overview}
\label{sect:motivation}



This section describes the role our novel IR in reshaping compiler design. Through illustrative examples, we demonstrate how the two optimization passes 
(i) minimize the 1D depth and (ii) the 2D size of transformed IR programs, while (iii) adhering to the constraint on temporal edge lengths.


At first glance, these optimization goals may seem to conflict with one another.
The first conflict arises between the 1D size and the lengths of temporal edges. While temporal edge lengths can be reduced by refreshing IR nodes after every fixed 1D depth~\cite{OnePerc}, this can increase the total 1D depth as each round of refresh incurs multiple additional 2D layers.
The second conflict involves the 2D size and the temporal edge lengths, as the reduction in 2D size in FlexLattice IR is enabled by its use of flexible temporal edges~\cite{OnePerc}.
Nonetheless, we show that these goals can be achieved simultaneously through the following two optimization passes.

\subsection{Dynamic Refreshing}
Our compiler is designed to be adaptive to a length limit of temporal edges by exploiting dynamic refreshing, which also leads to a reduction in 1D depth. We illustrate this with an example in Fig.~\ref{fig:overview_example}(c1)(c2), which shows a bounded limit of 4 for all temporal edges and a reduction in 1D depth from $6+m$ to $6$.

In Fig.~\ref{fig:overview_example}(c1)(c2), black edges represent computing edges, which are edges forming the measurement pattern used for computation, while orange edges represent refreshing edges, which are edges measured in fixed bases merely for node refresh. A node refresh can reduce storage time of the node in the delay line by remapping the node to the current 2D layer, where computation on that node can be continued after it is refreshed. For example, in Fig.~\ref{fig:overview_example}(c1), node $M$ is refreshed by refreshing edge $e2$ to a new node $M'$, which reduces the length of computing edge $e0$. After this refresh, the computation on node $M$ can be continued on $M'$ with computing edge $e7$.

In previous work \cite{OnePerc}, computation and refreshing are conducted alternately, as shown in Fig.~\ref{fig:overview_example}(c1). Every time after a certain depth of computation, all nodes stored in delay lines are refreshed by allocating additional multiple 2D layers for refresh only (the orange layers). Despite its prevention of very long temporal edges, the length limit is still unbounded since it is hard to control the length of each temporal edge. In particular, the number of required additional 2D layers $m$ varies in every round of refresh based on how many nodes need to be refreshed and how many of them have conflicted 2D coordinates. This induces an unpredictable overhead to the lengths of temporal edges. For example, the lengths of edge $e2,e4,e5$ and $e6$ are $2+m, 1+m, 2+m$ and $1+m$, all relying on the number of additional refreshing layers $m$.

In our compiler, each node is refreshed dynamically based on their need, without a distinction between computing layers and refreshing layers. On every 2D layer, nodes are refreshed selectively to prevent temporal edges from exceeding the length limit.
This is demonstrated in Fig.~\ref{fig:overview_example}(c2) where the length limit of temporal edges is set to be 4. Each edge $e'$ in Fig.~\ref{fig:overview_example}(c2) is a correspondence of edge $e$ in Fig.~\ref{fig:overview_example}(c1). 
This dynamic refreshing exhibits the following advantages.
\textbf{First}, it ensures a bounded length limit for temporal edges. In Fig.~\ref{fig:overview_example}(c1), edge $e1$ has a length $>4$ as its refresh has to wait until the beginning of the refreshing round. In contrast, in Fig.~\ref{fig:overview_example}(c2), the length of edge $e1'$ is reduced to 4, since refresh is enforced once the limit 4 is reached. Similarly, the lengths of edge $e2,e4,e6$ and $e6$ are also reduced to 4, no longer relying on an unbounded value $m$.
\textbf{Second}, the elimination of dedicated refreshing layers naturally reduces the 1D depth, as computation on the refreshed nodes do not need to wait for the beginning of the next computing round. In Fig.~\ref{fig:overview_example}(c2), the 1D depth is reduced from $6+m$ (Fig.~\ref{fig:overview_example}(c1)) to $6$. 


\subsection{2D-bounded Temporal Routing }
Our compiler exploits the broader optimization space enabled by the new IR with a 2D-bounded temporal routing, which can reduce the 2D size and 1D depth of IR programs. We demonstrate this with an example in Fig.~\ref{fig:overview_example}(d1)(d2), which shows a 2D size reduction from $3\times 3$ to $2\times 3$ and a 1D depth reduction from 4 to 3.

On layer 0, node $A,B,C,D$ and $E$ are located in $(0,0),(1,0),(1,1)$, $(0,1)$ and $(0,2)$, denoted by $A_0,B_0,C_0,D_0$ and $E_0$, with the subscripts indicating their 2D layers. To implement an MBQC program equivalent to a circuit shown in Fig.~\ref{fig:overview_example}(d3), we aim to exchange the 2D positions of $B$ and $C$ through subsequent layers so that $AC$, $BD$, $AD$ and $BC$ can all be connectable. In the previous FlexLattice IR, this exchange can be achieved as shown in Fig.~\ref{fig:overview_example}(d1), by the blue routing path between $B_0, B_2$ and the purple routing path between $C_0,C_2$, so that $AC$ and $BD$ are connected on layer 2 as $A_2 C_2$ and $B_2 D_2$ while $AD$ and $BC$ are connected on layer 3 as $A_3 D_3$ and $B_3 C_3$. This requires a 2D size of $3\times3$ and a 1D depth of 4. 

In our new IR where temporal edges are allowed to be skewed, this can be easily realized by skewing edges within a 2D distance, as shown by the blue skewed edge $B_0 B_1$ and the purple skewed edge $C_0 C_1$ in Fig.~\ref{fig:overview_example}(d2). This reduces the required 2D size to $2\times3$ and reduces the required 1D depth to 3. In this example, the skewed edges facilitate an easy swap of nodes, which resembles a native SWAP gate in the circuit model. However, in general cases, these temporal edges can be skewed in more ways to enable a flexible routing beyond swapping. That is why our 2D-bounded temporal routing is more powerful in reducing the 3D size of IR than simply incorporating a native SWAP into our baseline, as will be demonstrated later in the evaluation.

\section{Framework Design}
\label{sect:algorithm}


\subsection{Compilation Flow and Complexity}
\label{sect: algorithm_overview}


\begin{figure}[h!]
    \centering
    \includegraphics[width=\linewidth]{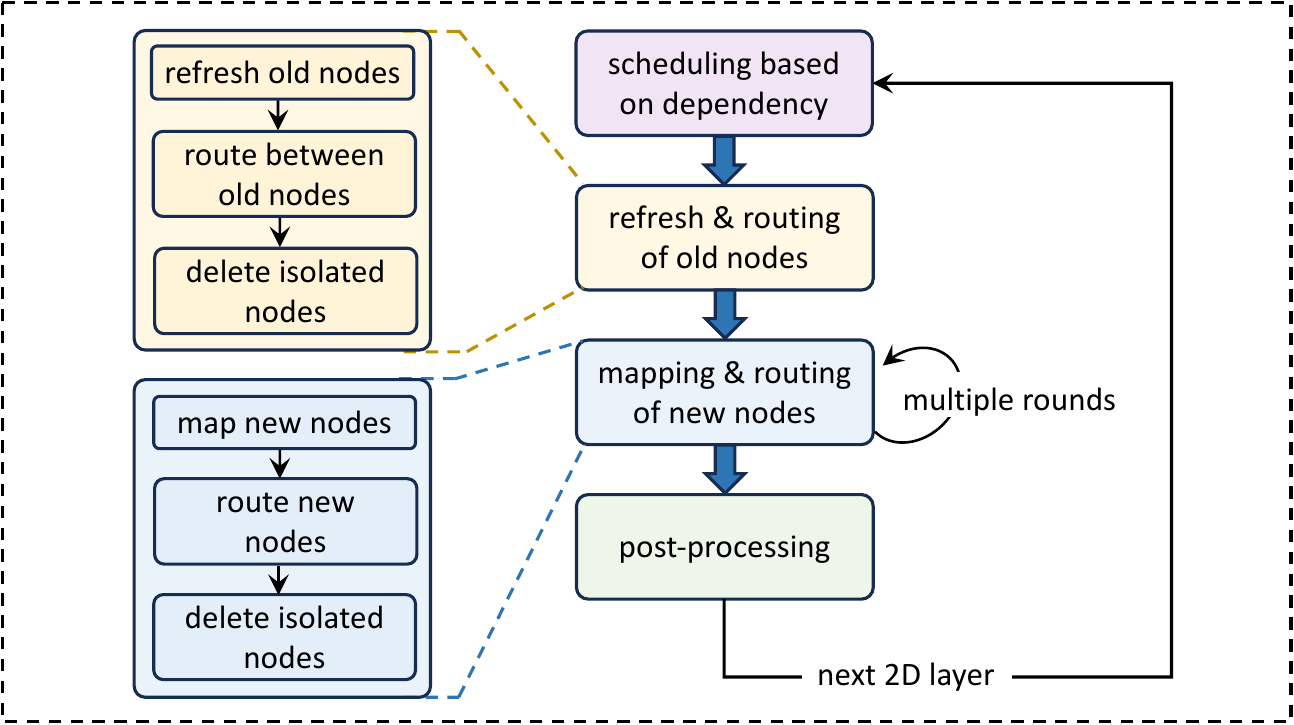}
    \caption{Compilation flow.}
    \label{fig:onemem_flow}
\end{figure}

\textbf{Scheduling} As shown in Fig.~\ref{fig:onemem_flow}, when given a MBQC program represented by a graph state, our compiler first schedules the nodes based on the dependencies among them. This is achieved by constructing a directed acyclic graph (DAG), with each directed edge indicating a dependency between graph state nodes.
To eliminate the need for a node to wait in delay lines before measurement, every time we only take the nodes in the front layer of the DAG, thus making sure that the node on the end of each edge is mapped to a later 2D layer than that of the beginning of the edge. 
 

\textbf{Refreshing old nodes} For each 2D layer, we first select some old nodes stored in delay lines and refresh them (yellow in Fig.~\ref{fig:onemem_flow}). The only exception is that, in the beginning of compilation when there are no nodes in the delay lines, we first map all input nodes in the graph state, allowing them to occupy multiple 2D layers if the given 2D size is small. Except for this initial stage, selected old nodes would be refreshed by mapping them to 2D positions near their original ones through 2D-bounded temporal routing. With the heap structure in our codes for node storage, the complexity of this step is $O(S\log S)$ where $S$ is the 2D size of IR.

\textbf{Routing old nodes} 
The next step is to map edges between these refreshed nodes (yellow in Fig.~\ref{fig:onemem_flow}). For each refreshed node, we check if their neighbors in the graph state are also refreshed. If so, we connect this refreshed node with its refreshed neighbors through shortest paths if allowed by the remaining space on the 2D layer. This routing is followed by a deletion of isolated nodes to save 2D space for subsequent steps. That is, if a refreshed node is not connected with any other node after this routing step, this isolated node will be removed from the 2D layer. The only exception is if it requires a mandatory refresh, which means that its storage time in the delay time is about to exceed the limit. With Dijkstra algorithm for shortest-path search, the complexity of this step is $O(S^3)$.

\textbf{Mapping and routing new nodes} After refreshing and connecting these old nodes, we then start mapping and routing new nodes (blue in Fig.~\ref{fig:onemem_flow}). Among nodes of the current front layer, every time from the graph state, we select a neighbor of a mapped node on the current 2D layer, mapping it to an adjacent position of that mapped node if allowed by the remaining 2D space. Then we route these new nodes to connect them through shortest paths if there are edges between them in the graph state. With Dijkstra algorithm for each shortest-path search, the complexity of this step is $O(S^3)$. 

\textbf{Multiple rounds of mapping and routing} This mapping and routing process is performed for multiple rounds if the 2D space remains (circular arrow in Fig.~\ref{fig:onemem_flow}). For each round after the first one, we first select some new nodes randomly from the current dependency layer, mapping them to random positions, and then select their neighbors in this dependency layer, mapping these neighbors to their adjacent positions. After that, we connect these new nodes through shortest-path routing if they have edges. Finally, if some nodes among the randomly selected nodes are isolated after this round, we remove them from the  2D layer to save space for subsequent rounds.

\textbf{Post-processing} After the refresh, mapping and routing on each 2D layer, a post-processing (green in Fig.~\ref{fig:onemem_flow}) will be conducted to deal with a special case where some nodes mapped on the current 2D layer have only one unmapped neighbor left in the graph state, which enables the space saving on subsequent 2D layers.

\subsection{Scheduling to Maintain Dependency Order}
\label{sect:scheduling}

\revise{Due to the non-determinism of quantum measurement outcomes, the measurement bases of some graph nodes are subject to real-time adjustments, depending on the measurement outcomes of other nodes.
These dependencies result in a partial order among measurements on different nodes, which can be determined by measurement calculus from previous work \cite{measurement_calculus, translation}.
As a preprocessing, we construct a DAG to represent the dependency relations between graph nodes. For each directed edge, the measurement basis of the node on the end depends on the measurement outcome of the node on the beginning.} 

\revise{The maintenance of dependency order in the compilation refers to mapping the node on the end of each directed edge to a later 2D layer than the one on the beginning, so that nodes do not need to wait for others after they are mapped. This is achieved by taking nodes in the front layer of the DAG in each round and dynamically updating the DAG during compilation.
Specifically, before refreshing and mapping nodes on each 2D layer, the DAG is updated by removing \emph{complete} nodes, where a node is considered as complete if all its edges are already mapped. Then, we take the front layer of the DAG and only schedule them in the subsequent mapping. The front layer consists of nodes without in-degrees in the updated DAG, which means all the nodes they depend on have become complete on previous 2D layers. In this way, we make sure that a node is always scheduled on a later 2D layer than nodes it depends on, thus removing the necessity for nodes to wait in delay lines before measurements.}



\subsection{Dynamic Refreshing}
\label{sect:dynamic_refreshing}

\begin{figure}[h!]
    \centering
    \includegraphics[width=\linewidth]{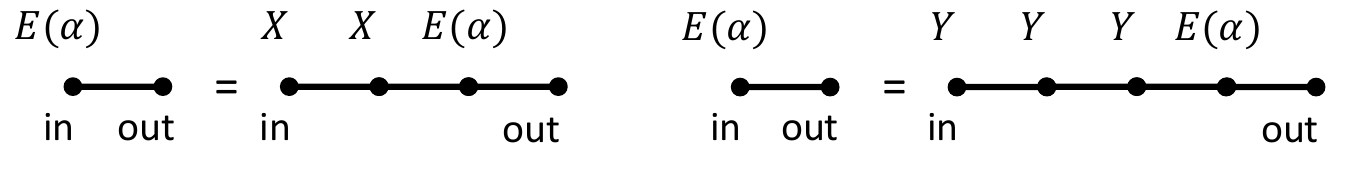}\\
    \hspace{0pt}(a)\hspace{100pt}(b)
    \caption{Insertion of identity patterns (a) XX and (b) YYY.}
    \label{fig:XX_YYY}
\end{figure}

\textbf{Node refresh}
The feasibility of node refresh lies in the fact that 
two consecutive X- or three consecutive Y-measurements can act as an identity operation that transfers an input state to another qubit up to some Pauli corrections \cite{mbqc2009,translation}.
, as illustrated in Fig.~\ref{fig:XX_YYY}. 
Hence when a node is about to exceed its storage time limit in delay lines, we can refresh it by mapping it onto the current layer again and inserting the identity operation.
Note that while a refresh needs two additional X- or three additional Y-measurements, we do not need to remap the node twice or three times, \revise{because each edge in the IR
is realized by a connected path of successful fusions on the physical hardware (Fig.~\ref{fig:fbqc_hardware}), which already consists of multiple photonic qubits. The choice of consecutive X- and Y-measurements along the path is determined by whether the path has an even (XX$\cdots$XX) or odd (XX$\cdots$YYY) length.}


\textbf{Selection of Nodes for Refresh} Before mapping graph state nodes on each 2D layer, we select and refresh some nodes stored in delay lines. The selection is prioritized from mandatory refresh to optional refresh, with the optional ones further prioritized based on their neighborhoods and storage time in delay lines.

\begin{figure}[h!]
    \centering
    \includegraphics[width=0.45\linewidth, height=0.33\linewidth]
    {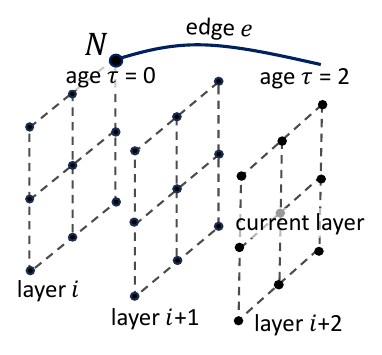}
    \caption{Age of an incomplete node.}
    
    \label{fig:node_age}
\end{figure}

According to the last layer on which it was mapped, each incomplete node can be labeled by an age $\tau$.
For example, node $N$ in Fig.~\ref{fig:node_age} is an incomplete node with an edge $e$ that is not yet mapped. Since the last time node $N$ was mapped is layer $i$, while the current layer is $i+2$, the age of node $N$ is $\tau=2$. When this age reaches the length limit for time-like edges, this node requires a mandatory refresh, otherwise its refresh is optional.

After collecting all nodes for mandatory refresh, 
then
some nodes will be selected for optional refresh if the 2D space allows. We select these nodes in the following order, with the nodes within each category selected in the descending order of their ages:
\begin{itemize}
    \item nodes having edges with mandatory refreshed nodes
    \item nodes having edges with other nodes in delay lines
    \item nodes only having edges with nodes not mapped yet
\end{itemize}


That is, we first select the nodes that are neighbors of the mandatory refreshed nodes in the graph state. Since the mandatory refreshed nodes have already been mapped, mapping the nodes connected to them enables their edges to be mapped via routing. Next, we select the nodes that have graph state neighbors in the delay lines. This allows us to map both connected nodes together, enabling their edges to be mapped through routing. Lastly, the nodes that only have edges with unmapped nodes would be mapped if allowed by the 2D space.



\textbf{Refresh Percentage Tuning} 
To reserve enough 2D space for new graph nodes, we restrict the percentage of positions on each 2D layer used for refresh by a refreshing bound $b_r$. This bound can be violated by mandatory node refresh, but not by optional refresh. That is, when the number of mandatory refreshed nodes $m<b_r$, at most $b_r-m$ nodes can be selected for optional refresh.

\revise{In our compiler, this refreshing bound $b_r$ is dynamically adjusted based on the size of 2D layers $S$ and the number of mandatory refreshed nodes $m_\mathrm{last}$ on the last 2D layer, in the following way:
\begin{equation}
b_r = \min\{m_\mathrm{last}/S + p, \enspace 1\}
\label{parameter_p}
\end{equation}
where p is an adjustable parameter with a default value of 0.4. 
The insight behind this is when the mandatory refreshed nodes on the last layer are too many, it implies that we need to refresh the old nodes more aggressively to prevent their accumulation in delay lines.
Note that a reduction of the parameter $p$ can help reduce the real-time overhead of path searching for time-like edges, at the cost of an increased 1D depth.}

\subsection{2D-bounded Temporal Routing}
\label{sect:braiding_style_routing}

\textbf{Inter-layer Routing} 
According to the 2D position a node was mapped last time, 
each node $v$ stored in delay lines is labeled by a 2D coordinate $(x_v,y_v)$.
When a node is refreshed on the current layer, it can be mapped around $(x_v,y_v)$ within a range allowed by skewed temporal edges. Similarly, when a new node in the neighborhood of $v$ is mapped onto the current layer, it can also be mapped around $(x_v,y_v)$ within this range. To reduce the overhead on physical hardware, our IR restricts this range to a Hamming distance of 1.
This is because starting from $\sqrt{2}$, an increased distance would lead to an increased resource overhead, as will be shown in Fig.~\ref{fig:IR_extention_eval}. 
This means that the refreshed nodes can be mapped to\[
(x', y')\in \mathrm{near}((x_v,y_v))\equiv (x_v, y_v)\cup\mathrm{adj}((x_v,y_v))
\] on the current 2D layer, where 
\begin{align*}
    \mathrm{adj}((x_v,y_v))\equiv \{&(x_v+1, y_v),(x_v-1, y_v),\\
    &(x_v, y_v+1),(x_v, y_v-1)\}
\end{align*}

\textbf{Routing Heuristics}
We now define the heuristics of choosing 2D positions for 2D-bounded temporal routing during node refresh. 
The choice of these 2D positions on the current layer aims to either minimize the distances between connected nodes in the graph state or maximize the available spaces around nodes. Among the selected nodes to refresh $R$, a node may or may not have connections with other selected nodes. For the connected ones, we minimize their distances by mapping each node $v$  to a position $(x',y')$ according to the following heuristic
\[
\min_{(x',y')\in \mathrm{near}((x_v,y_v))}\sum_{n\in R\cap\mathrm{neigh}(v)}\mathrm{dist}((x',y'), (x_n,y_n))
\]
where $\mathrm{neigh}(v)$ stands for the neighbors of $v$ in the graph state, and the coordinate $(x_n,y_n)$ stands for the position each neighboring node $n$ in the memory. For the nodes with no connections in $R$, we map each node $u$ to a position according to the following heuristic
\[
\max_{(x',y')\in \mathrm{near}((x_u,y_u))}\sum_{n\in\mathrm{near}((x',y'))}\Theta(\text{$n$ not occupied})
\]
where $\Theta$ is the step function with a value 1 when the condition is satisfied and a value 0 otherwise.

\textbf{Post-processing} In the steps above, each new node can either connect to a refreshed node or another new node, but can not connect directly with old nodes stored in delay lines. In the post-processing of each 2D layer, we enable this opportunity by addressing a special case. That is, when a node mapped on the current 2D layer is found to have only one unmapped neighbor left in the graph state, then next time when we refresh this node, we will replace the refreshed node with its unmapped neighbor directly.


\subsection{Realization of Extended IR}
\label{sect:implementation}


On physical hardware, edges of an IR program are realized through a (2+1)-D path searching among successful fusions, as illustrated earlier in Fig.~\ref{fig:fbqc_hardware}(c)(d).
The realization of skewed edges in the IR requires little modification on the physical level. The only thing we need to do is allow path searching between qubits corresponding to IR nodes at \textbf{nearby} 2D locations of different 2D layers. Intuitively, this more flexible path searching can increase resource overhead by increasing the PL ratio, as the paths for different skewed edges in various directions can more easily conflict with each other. 
However, through simulation, we find that this overhead is negligible when the 2D range of the skewed edges is restricted to only 1. 
\section{Evaluation}\label{sect:eval}

\subsection{Experiment Setup}
\label{sect:setup}
\textbf{Metrics} 
We evaluate the performance of our compiler with the following four metrics of transformed IR programs: 1D depth, 2D size, the maximal length of temporal edges $D_f$ (i.e., the maximal wait time for fusion), and the maximal wait time for measurement. We only explain the last one here as the other three have been explained in earlier sections. Besides the unbounded number of fusion devices, OnePerc \cite{OnePerc} also requires an unbounded number of measurement devices due to its over-flexible scheduling, which
can cause an IR node to wait excessively long in delay lines before
measurement. 
Hence we list this wait time for measurement required by compiled programs as the fourth metric.


\textbf{Benchmark programs} 
We select a set of benchmark programs for NISQ and FT respectively. Benchmarks for NISQ include Quantum Approximate Optimization Algorithm (QAOA), Ripple-Carry Adder (RCA)~\cite{RCA} and Variational Quantum Eigensolver (VQE). 
QAOA is implemented on randomly generated 3-regular graphs, where ZZ gates are applied only between graph-connected qubits.
For VQE, we follow the commonly used full-entanglement ansatz, which proves to be an expressive ansatz \cite{qiskit_vqe, vqe_ansatz}. 
Benchmarks for FT include Quantum Fourier transform (QFT), Grover's algorithm (Grover) \cite{grover} and quantum simulation (QSIM) \cite{qsim0, qsim1}. For QSIM, we use XXZ Heisenberg Hamiltonians as the target model and apply Trotterization to generate the circuit.
In the evaluation, each `benchmark-$n$' stands for a program corresponding to an $n$-qubit circuit in the circuit model.

\textbf{Hardware Settings} The best known scheme for fusion implementation has a theoretical success rate of 78\% \cite{0.78}, whereas a success rate of 75\% is achievable with significantly less resource overhead \cite{0.78,0.95with30}. As a result, 
we adopt the 75\% fusion success rate in the experiments which demonstrate the feasibility of IR extension, and provide a sensitivity analysis by varying it from 66\% to 78\%. By fixing the PL ratio as 4 (i.e., a slightly higher value than the natural PL ratio of $\sim 3.1$ to ensure success of IR realization), our IR implementation requires $D_f+1$ fusion devices with connected delay lines on each chiplet, with the longest storage time of photons being $4D_f$ RSLs. With state-of-the-art fiber technology with a loss rate of 0.2dB/km \cite{fiber}, photons can be stored for 1500 RSLs with a $\sim 5\%$ loss, given an RSG clock cycle $\sim 1ns$ \cite{interleaving}. With our largest \todoo{$D_f=10$} in the evaluation, the photon loss rate after stored for $4D_f$ is hence as small as \todoo{0.12\%}. 



\textbf{Baseline 1} We compare our framework with state-of-the-art photonic MBQC compiler \textbf{OnePerc} \cite{OnePerc}. 
Since OnePerc is unable to bound lengths of temporal edges and wait time for measurements, we adopt the following strategies to make it more comparable with our compiler, at the cost of increasing their 1D depth to some extent. First, when a $D_f$ is specified, we ask OnePerc to perform its periodic refresh after every $D_f$ layers and record the longest temporal edge.
Second, we reduce its measurement wait time by posing a more strict scheduling. That is, 
we restrict the number of dependency layers that can appear on each layer (2 is achievable) and record the wait time limit for measurement under this restriction. 
Note that while these strategies can mitigate the problems, they can not guarantee an adaptive bound for these limits.



\textbf{Baseline 2} Since OnePerc is not capable of bounding temporal edge lengths and measurement wait time, we compare with a secondary baseline with this capability, to demonstrate the performance of our compiler under such restrictions. This baseline is adapted from \textbf{Qiskit} \cite{qiskit}, state-of-the-art compiler in the circuit model. The feasibility of adapting a circuit model compiler to an MBQC compiler lies in the correspondence between the space-time diagram of a circuit and the structure of cluster states (i.e., $D_f=1$) \cite{mbqc2009}. 
This baseline guarantees a length limit of 1 for temporal edges and a wait time limit of 0 for measurements. However, it requires a minimal 2D size of $n$ for each $n$-qubit benchmark and results in a longer 1D depth.

\subsection{Feasibility of IR Extension}
\begin{figure*}[tp!]
    \centering
    \includegraphics[width=0.33\linewidth]{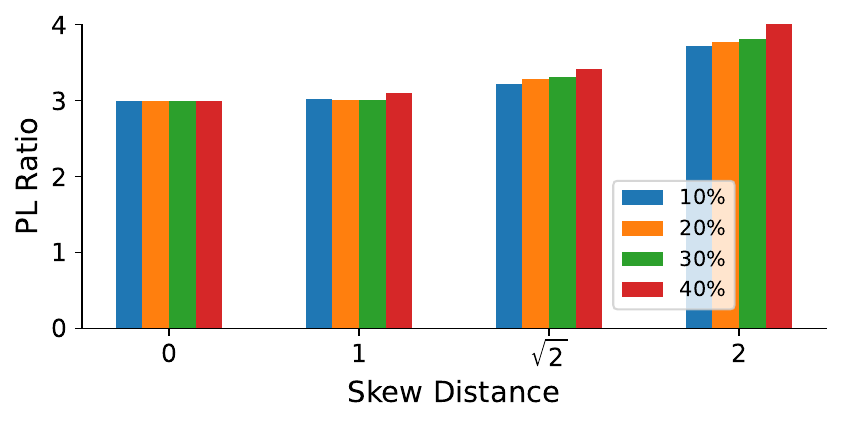}
    \includegraphics[width=0.33\linewidth]{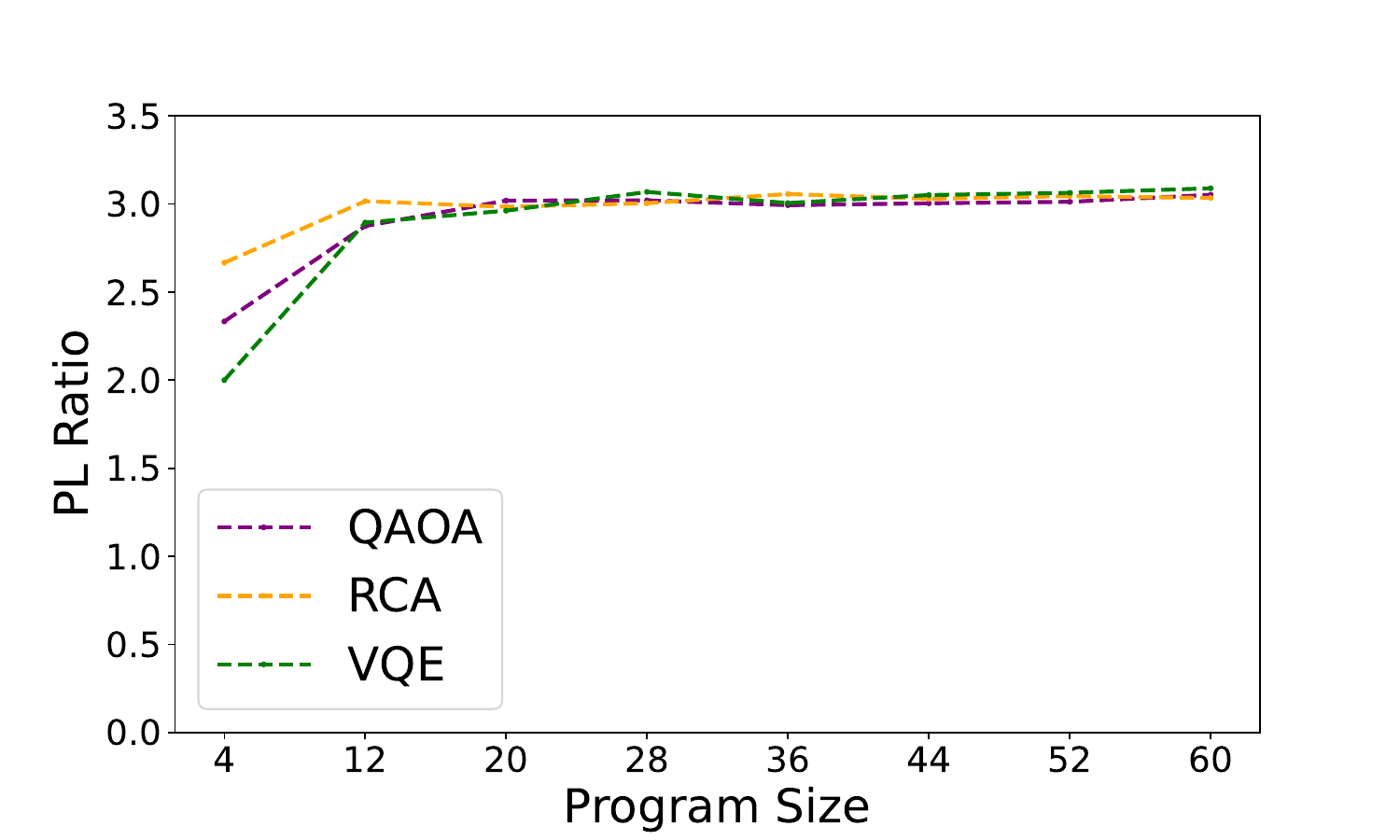}
    \includegraphics[width=0.33\linewidth]{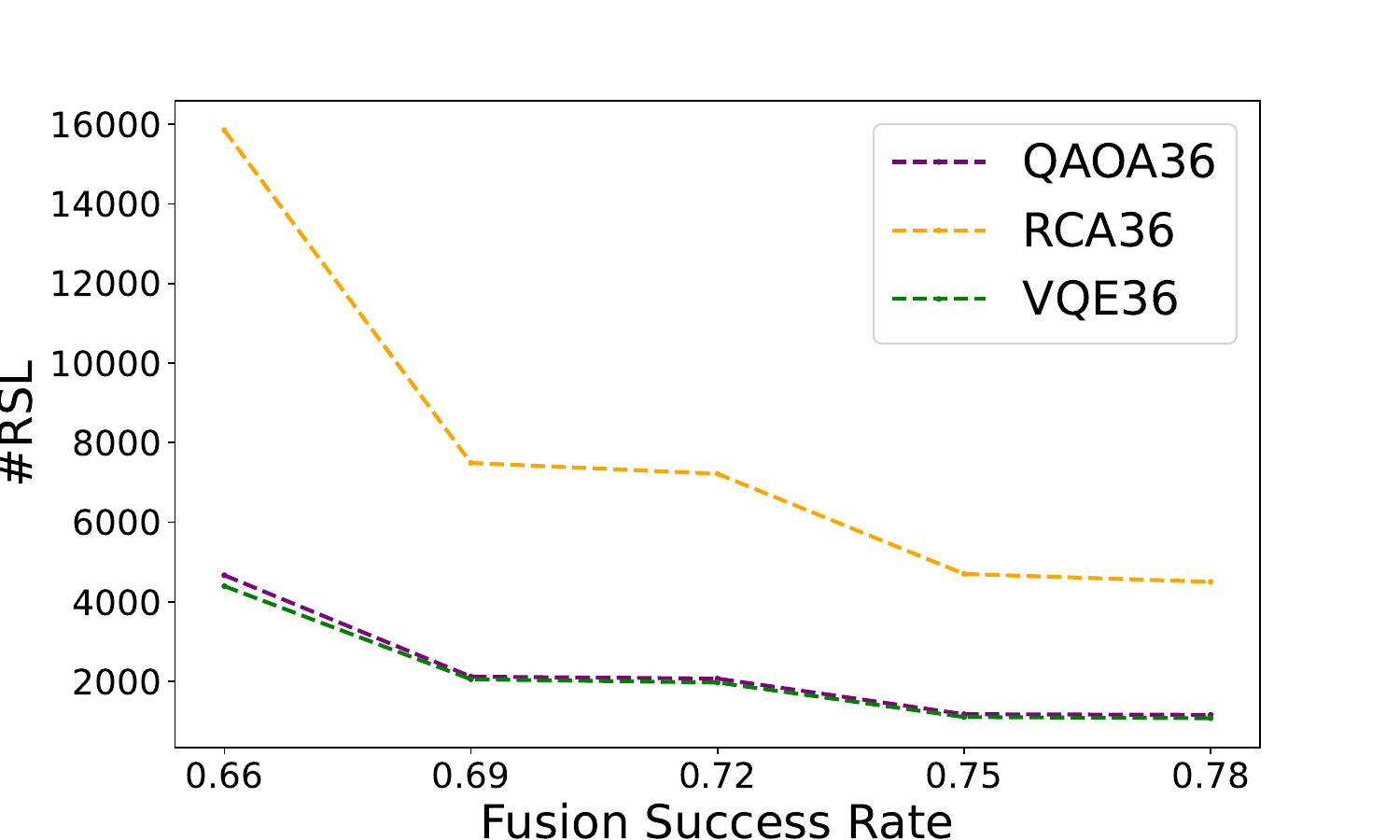}
    \caption{Effects of IR extension.}
    \label{fig:IR_extention_eval}
\end{figure*}

\textbf{Negligible Effect on PL Ratio}
Fig.~\ref{fig:IR_extention_eval}(a) shows the PL ratio as the allowed 2D distance for skewed edges increases. As a preliminary estimation of this overhead, we take an extended IR with a \todoo{$6\times6$} 2D size, randomly selecting \todoo{10\% - 40\%} of all possible skewed edges between 2D layers, skewing them in random directions for a random 2D distance within the allowed limit. It can be seen that the PL ratio increases with the allowed skew distance, and increases slightly as temporal edges become denser. Moreover, when the skew distance is restricted to only 1, the PL ratio is about the same as that of FlexLattice IR, which corresponds to a 2D skew limit of 0. 

To prevent an increased overhead, our extended IR restricts the 2D distance of skewed edges to only 1. To illustrate its overhead for real programs, we further compile our benchmarks to the extended IR with a 2D size of \todoo{$6\times6$}.
Fig.~\ref{fig:IR_extention_eval}(b) illustrates the variation of PL ratio as the program size increases. 
It can be seen that the PL ratio first increases with the program size and then becomes stable at a value $\sim 3.1$. This aligns with the result of PL ratio $\sim 3.1$ for FlexLattice IR, implying that when restricting the skew distance to 1, the introduction of skewed edges in our new IR incurs a negligible hardware overhead.

\textbf{Negligible Effect on Fusion Failure Tolerance} Our extended IR implementation does not reduce the tolerance for fusion failures either when compared to OnePerc \cite{OnePerc}. Fig.~\ref{fig:IR_extention_eval}(c) shows the number of RSLs when executing 36-qubit programs 
on RSLs of \todoo{$84\times84$} 2D size under different fusion success probabilities. It can be seen that similar to OnePerc, our compiler can tolerate a fusion success probability as low as 0.66, with the number of RSLs decreasing with a higher fusion success probability. This is consistent with the performance of OnePerc.

\begin{table*}[tp]
    \centering
    \caption{Results of our compiler and the baseline. Each `benchmark-$n$' corresponds to an $n$-qubit circuit in the circuit model.
    }
    \resizebox{\textwidth}{!}{
        \renewcommand*{\arraystretch}{1.05}
        \begin{normalsize}
            \begin{tabular}{|p{2.3cm}|p{1.8cm}|p{1.4cm}|p{1.4cm}|p{1.4cm}|p{1.7cm}|p{1.4cm}|p{1.6cm}|p{2.4cm}|p{2.1cm}|}
        
        \hline
        Target Time-like Edge Length Limit $D_f$ (layers)  & Benchmark \ \ \ \ \ \ \ \ \ \   - \#Qubit & OnePerc \ \ \ \ \ \ \ \ 1D Depth & Our \ \ \ \ \ \ \ \ \ \ 1D Depth & Depth \ \ \ \ Improv.  & OnePerc $D_f$ \ \ \ \ (compiled) & Our $D_f\quad$ \ \ \ \ (compiled) & $D_f$ Improv. & OnePerc Measurement Wait Time $\quad$  (layers)  & Our Measurement Wait Time (layers)\\
        \hline
        \multicolumn{1}{|c|}{\multirow{9}{*}{\parbox{2cm}{ 30 virtual layers}}}
        & QAOA-36 & 1,020 & 235 & \textbf{ 4.34 $\times$} & 44 & 30 & \textbf{ 1.47 $\times$} & 42 & \textbf{ 0 }\\
        \cline{2-10}
        & RCA-36 & 1,518 & 921 & \textbf{ 1.65 $\times$} & 39 & 30 & \textbf{ 1.3 $\times$} & 37 & \textbf{ 0 }\\
        \cline{2-10}
        & VQE-36 & 803 & 210 & \textbf{ 3.82 $\times$} & 43 & 21 & \textbf{ 2.05 $\times$} & 139 & \textbf{ 0 }\\
        \cline{2-10}
        & QAOA-64 & 2,267 & 605 & \textbf{ 3.75 $\times$} & 53 & 30 & \textbf{ 1.77 $\times$} & 51 & \textbf{ 0 }\\
        \cline{2-10}
        & RCA-64 & 2,180 & 1,777 & \textbf{ 1.23 $\times$} & 38 & 30 & \textbf{ 1.27 $\times$} & 34 & \textbf{ 0 }\\
        \cline{2-10}
        & VQE-64 & 1,627 & 712 & \textbf{ 2.29 $\times$} & 49 & 30 & \textbf{ 1.63 $\times$} & 637 & \textbf{ 0 }\\
        \cline{2-10}
        & QAOA-100 & 5,902 & 1,059 & \textbf{ 5.57 $\times$} & 63 & 30 & \textbf{ 2.1 $\times$} & 61 & \textbf{ 0 }\\
        \cline{2-10}
        & RCA-100 & 3,846 & 2,516 & \textbf{ 1.53 $\times$} & 40 & 30 & \textbf{ 1.33 $\times$} & 36 & \textbf{ 0 }\\
        \cline{2-10}
        & VQE-100 & 3,150 & 1,301 & \textbf{ 2.42 $\times$} & 54 & 30 & \textbf{ 1.8 $\times$} & 1,391 & \textbf{ 0 }\\
        \hline
        \multicolumn{1}{|c|}{\multirow{9}{*}{\parbox{2cm}{ 20 virtual layers}}}
        & QAOA-36 & 1,073 & 207 & \textbf{ 5.18 $\times$} & 35 & 10 & \textbf{ 3.5 $\times$} & 31 & \textbf{ 0 }\\
        \cline{2-10}
        & RCA-36 & 1,799 & 837 & \textbf{ 2.15 $\times$} & 28 & 20 & \textbf{ 1.4 $\times$} & 25 & \textbf{ 0 }\\
        \cline{2-10}
        & VQE-36 & 866 & 224 & \textbf{ 3.87 $\times$} & 31 & 20 & \textbf{ 1.55 $\times$} & 109 & \textbf{ 0 }\\
        \cline{2-10}
        & QAOA-64 & 2,472 & 594 & \textbf{ 4.16 $\times$} & 44 & 20 & \textbf{ 2.2 $\times$} & 40 & \textbf{ 0 }\\
        \cline{2-10}
        & RCA-64 & 2,063 & 1,615 & \textbf{ 1.28 $\times$} & 28 & 20 & \textbf{ 1.4 $\times$} & 25 & \textbf{ 0 }\\
        \cline{2-10}
        & VQE-64 & 1,875 & 620 & \textbf{ 3.02 $\times$} & 39 & 20 & \textbf{ 1.95 $\times$} & 716 & \textbf{ 0 }\\
        \cline{2-10}
        & QAOA-100 & 7,361 & 1,069 & \textbf{ 6.89 $\times$} & 52 & 20 & \textbf{ 2.6 $\times$} & 51 & \textbf{ 0 }\\
        \cline{2-10}
        & RCA-100 & 4,049 & 2,485 & \textbf{ 1.63 $\times$} & 28 & 20 & \textbf{ 1.4 $\times$} & 26 & \textbf{ 0 }\\
        \cline{2-10}
        & VQE-100 & 3,686 & 1,268 & \textbf{ 2.91 $\times$} & 40 & 20 & \textbf{ 2.0 $\times$} & 1,868 & \textbf{ 0 }\\
        \hline
        \multicolumn{1}{|c|}{\multirow{9}{*}{\parbox{2cm}{ 10 virtual layers}}}
        & QAOA-36 & 1,502 & 202 & \textbf{ 7.44 $\times$} & 22 & 10 & \textbf{ 2.2 $\times$} & 21 & \textbf{ 0 }\\
        \cline{2-10}
        & RCA-36 & 1,912 & 745 & \textbf{ 2.57 $\times$} & 17 & 10 & \textbf{ 1.7 $\times$} & 16 & \textbf{ 0 }\\
        \cline{2-10}
        & VQE-36 & 1,087 & 194 & \textbf{ 5.6 $\times$} & 21 & 10 & \textbf{ 2.1 $\times$} & 20 & \textbf{ 0 }\\
        \cline{2-10}
        & QAOA-64 & 3,818 & 530 & \textbf{ 7.2 $\times$} & 34 & 10 & \textbf{ 3.4 $\times$} & 32 & \textbf{ 0 }\\
        \cline{2-10}
        & RCA-64 & 2,448 & 1,430 & \textbf{ 1.71 $\times$} & 18 & 10 & \textbf{ 1.8 $\times$} & 16 & \textbf{ 0 }\\
        \cline{2-10}
        & VQE-64 & 2,617 & 609 & \textbf{ 4.3 $\times$} & 30 & 10 & \textbf{ 3.0 $\times$} & 457 & \textbf{ 0 }\\
        \cline{2-10}
        & QAOA-100 & 9,825 & 930 & \textbf{ 10.56 $\times$} & 40 & 10 & \textbf{ 4.0 $\times$} & 57 & \textbf{ 0 }\\
        \cline{2-10}
        & RCA-100 & 4,977 & 2,389 & \textbf{ 2.08 $\times$} & 18 & 10 & \textbf{ 1.8 $\times$} & 16 & \textbf{ 0 }\\
        \cline{2-10}
        & VQE-100 & 5,745 & 1,094 & \textbf{ 5.25 $\times$} & 34 & 10 & \textbf{ 3.4 $\times$} & 3,286 & \textbf{ 0 }\\
        \hline
    \end{tabular}
    \end{normalsize}
    }
    \label{tab:main_table}
\end{table*}

\subsection{Experiment Results}

This subsection shows our experiment results of the transformed IR programs. Due to the natural tradeoff between the 2D size and 1D depth \cite{OnePerc}, we first compare the last three metrics with the baselines while fixing the 2D size to a \emph{standard 2D size} of $\sqrt{n}\times\sqrt{n}$ for each $n$-qubit program.
This 2D size is then varied in subsequent experiments to demonstrate the adaptability of our compiler to various hardware sizes. After that, we perform an ablation study by enabling and disabling the 2D-bounded temporal routing.

\textbf{Compared to Baseline 1}
Table~\ref{tab:main_table} presents the comparison of our framework with OnePerc as the specified length limit $D_f$ of temporal edges and the program size vary, while the 2D size of IR is fixed to the standard size. The improvement factors are obtained from the ratios between the baseline and our compiler. The results show that our compiler achieves a significant reduction in 1D depth. The average improvement factor of \todoo{3.68} becomes increasingly significant as \( D_f \) becomes more restricted.
 Moreover, the lengths of temporal edges in our compiler are all within the specified $D_f$, which demonstrates its ability of restricting the required number of fusion devices to an adaptive limit. In contrast, OnePerc has an average excess of \todoo{89.9\%} over the specified limit, \todoo{with the excess increasing as the size of programs increases and as the specified $D_f$ decreases}. On average, the improvement factor in the length limit is \todoo{2.28}. Furthermore, our compiler also successfully reduces the wait time for measurement to 0, thus requiring only one measurement device on each chiplet. In contrast, OnePerc has an average maximal wait time of \todoo{264} layers, with the maximal wait time being $>1000$ occasionally if node congestion occurs.



\textbf{Compared to Baseline 2}
When the length limit of temporal edges is restricted to $D_f=1$, Table~\ref{tab:secondary_table} presents the comparison of our framework with the secondary baseline adapted from Qiskit as described in Section~\ref{sect:setup}. It can be seen that our compiler significantly reduces the 1D depth when the 2D size of IR is fixed to the standard size, with an average improvement factor of \todoo{3.56}. We have omitted the wait time for measurement in Table~\ref{tab:secondary_table}, as both the baseline and our compiler have a wait time of 0.


\begin{table}[h!]
    \centering
    \caption{Results of our compiler when $D_f=1$ and the baseline. 
    }
    \resizebox{\linewidth}{!}{
        \renewcommand*{\arraystretch}{1}
        \begin{normalsize}
            \begin{tabular}{|p{1.4cm}|p{0.9cm}|p{0.9cm}|p{0.7cm}|p{1.2cm}|p{1.6cm}|}
        
        \hline
        Benchmark - \#Qubit  & Qiskit Depth & Qiskit + Skew Depth & Our Depth & Depth Improv.  vs Qiskit  & Depth $\quad$ Improv. vs Qiskit+Skew\\ \hline
        QAOA-36 & 1,055 & 739 & 264 & 4.0 & 2.8 \\\hline
        RCA-36 & 2,379 & 1,660 & 561 & 4.24 & 2.96 \\\hline
        VQE-36 & 830 & 367 & 191 & 4.35 & 1.92 \\\hline
        QAOA-64 & 3,279 & 1,810 & 557 & 5.89 & 3.25 \\\hline
        RCA-64 & 4,397 & 2,994 & 1,075 & 4.09 & 2.79 \\\hline
        VQE-64 & 1,929 & 914 & 452 & 4.27 & 2.02 \\\hline
        QAOA-100 & 6,089 & 3,297 & 1,048 & 5.81 & 3.15 \\\hline
        RCA-100 & 6,896 & 4,806 & 1,595 & 4.32 & 3.01 \\\hline
        VQE-100 & 4,723 & 1,780 & 813 & 5.81 & 2.19 \\\hline
   
    \end{tabular}
    \end{normalsize}
    }
    \label{tab:secondary_table}
\end{table}

        
                

\begin{figure*}[tp!]
        \centering
        \includegraphics[width=0.33\linewidth]{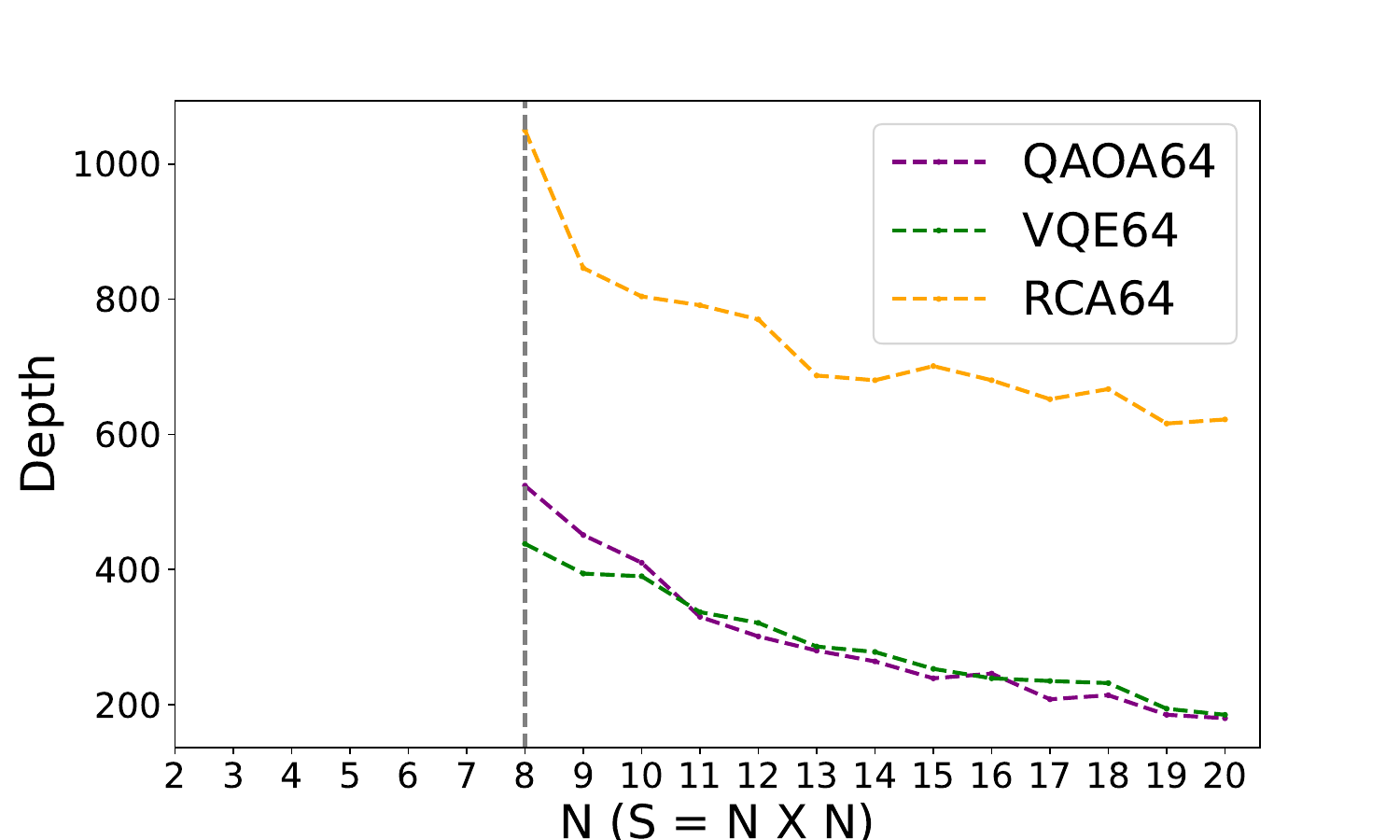}
        \includegraphics[width=0.33\linewidth]{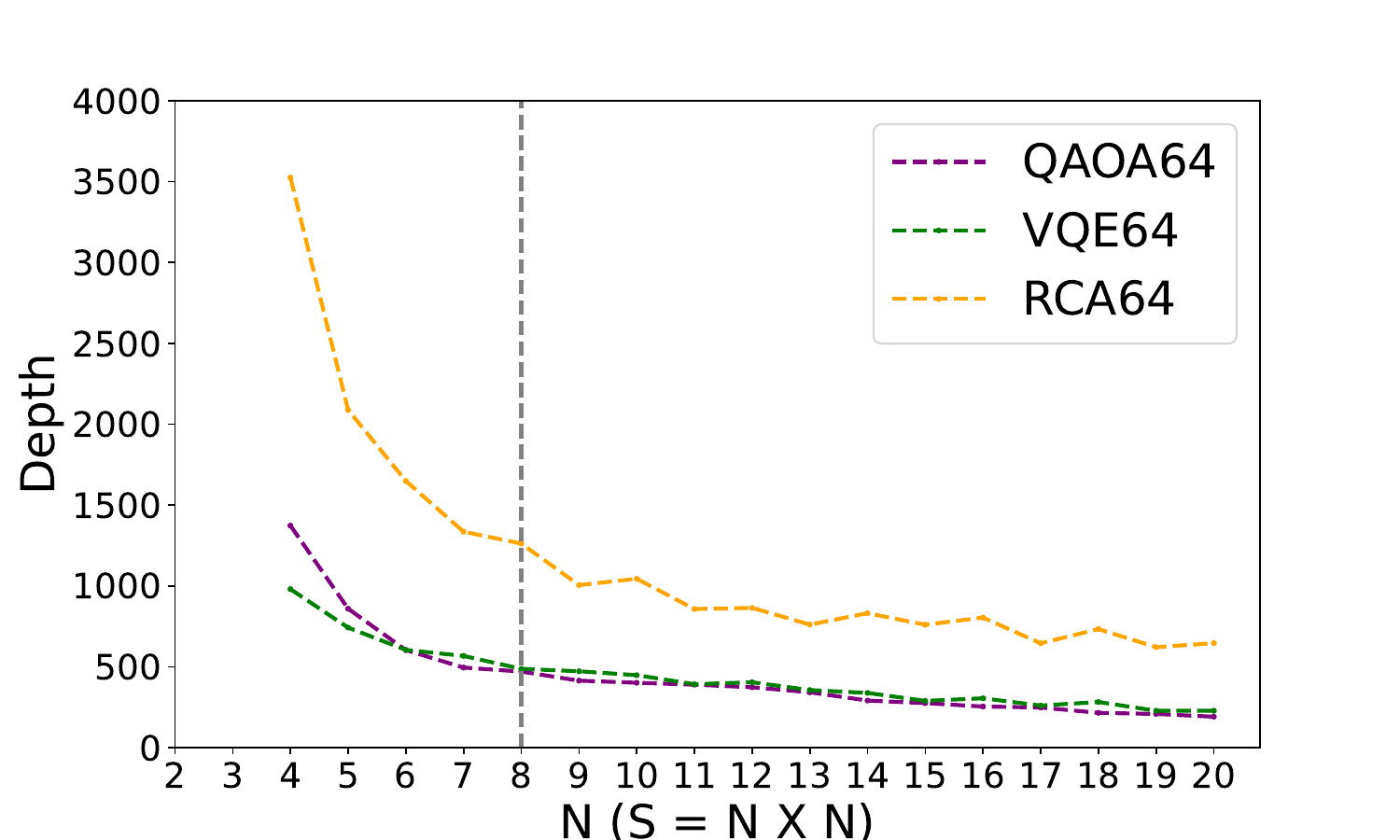}
        \includegraphics[width=0.33\linewidth]{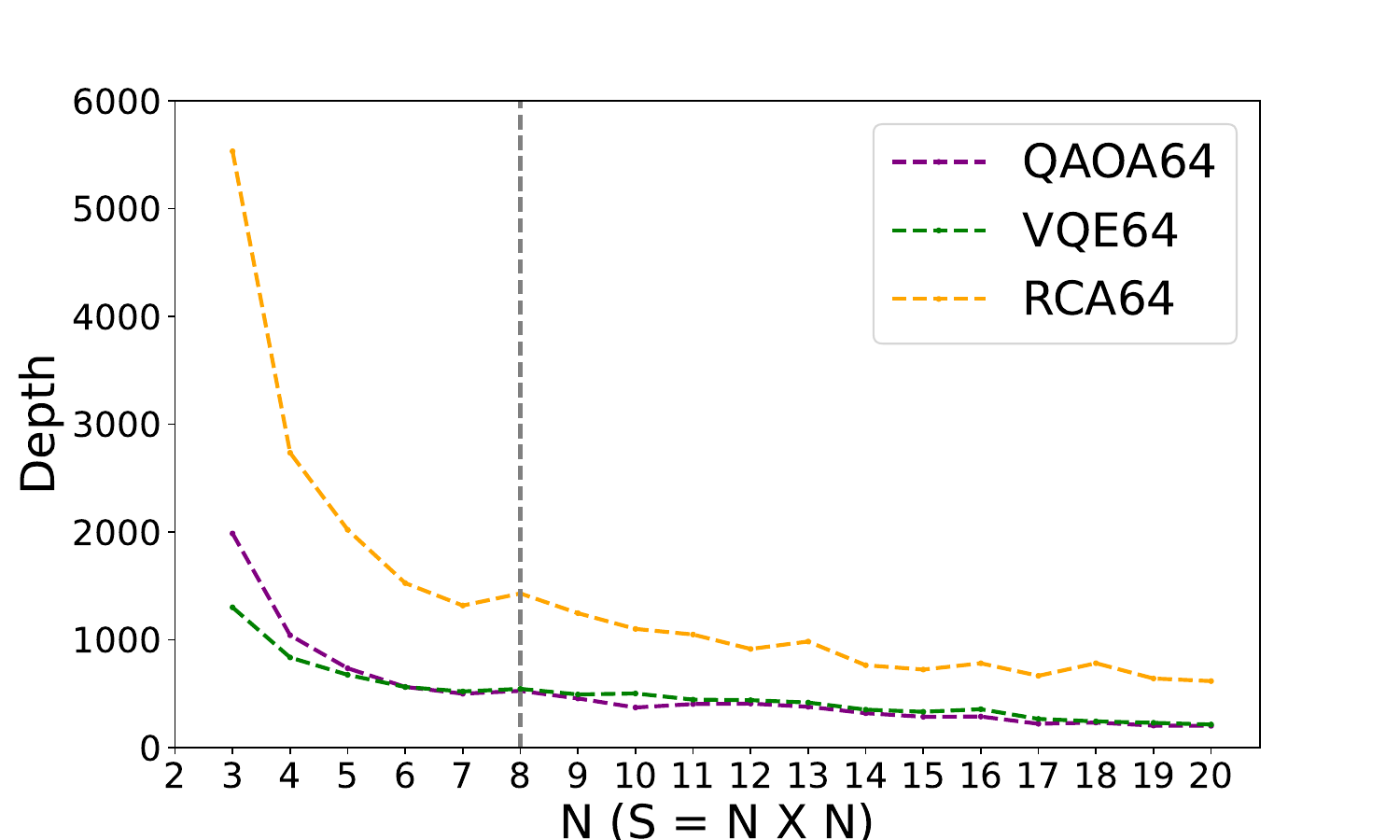}
        \hspace{10pt}(a)\hspace{160pt}(b)\hspace{160pt}(c)
        \caption{Effects of 2D size on 64-qubit programs when $D_f=1$ (a), $D_f=5$ (b) and \todoo{$D_f=10$} (c). Dashed lines correspond to the standard 2D size, which is $S=8\times8$ for 64-qubit programs.}
        \label{fig:virtual_hardware_size_sensitivity}
\end{figure*}

We further compare our compiler with a strengthened version of baseline 2. That is, we allow native SWAP gates in baseline 2 to reduce its overall 1D depth, with the native SWAP gates implemented by our 2D-bounded temporal routing.
As shown in Table~\ref{tab:secondary_table}, our compiler also achieves a significantly smaller 1D depth than this strengthened baseline 2, with an average improvement factor of \todoo{2.01}. This is because our 2D-bounded temporal routing is more flexible than a native SWAP.
In the routing process, our compiler tries to find the best 2D position for each node considering the overall mapping, not limited to swapping the 2D positions of different nodes.

\textbf{Reduction in 2D size} Our framework can adapt to different 2D sizes of IR, with a tradeoff in 1D depth. A reduced 2D size indicates a reduced requirement for the number of chiplets and a reduced latency for real-time processing that handles fusion failures. This adaptability aligns more closely with the OnePerc baseline, standing in contrast to the Qiskit baseline, which can only utilize a fixed hardware size — it can neither operate on a smaller hardware nor effectively leverage a larger hardware.


The adaptability of our compiler to a smaller 2D size relies on the allowance for longer temporal edges.
Fig.~\ref{fig:virtual_hardware_size_sensitivity}(a)(b)(c) illustrates the variation of 1D depth with 2D size $S$ for 64-qubit programs when the length limit $D_f$ of temporal edges is \todoo{$1, 5$ and 10}, respectively. This is obtained by compiling each program to an increasingly smaller 2D size for each $D_f$, until the compilation gets stuck by node congestion. The vertical dashed lines in Fig.~\ref{fig:virtual_hardware_size_sensitivity} corresponds to the standard 2D size (\todoo{$8\times8$}) used in previous experiments. It can be seen that the smallest 2D size required by our compiler decreases with an increased length limit for temporal edges. When \todoo{$D_f=10$}, the 2D size of these 64-qubit programs can be reduced to as small as $3\times3$, which is $7\times$ smaller than the standard size of $8\times8$.

Moreover, Fig.~\ref{fig:virtual_hardware_size_sensitivity} exhibits the tradeoff between 2D size and 1D depth, with the program depth growing with a reduced 2D size. 
Despite this rapid growth in 1D depth, the 3D volume, defined as 2D size $\times$ 1D depth, still decreases. For example, in Fig.~\ref{fig:virtual_hardware_size_sensitivity}, the 3D volume is \todoo{26,457} when 2D size is $3\times3$, which is \todoo{50.4\%} less than the 3D volume of \todoo{53,355} when 2D size is $8\times8$.

\textbf{\revise{Effect of 2D-bounded Temporal Routing.}}
Table~\ref{tab:secondary_table} has shown the effect of allowing skewed edges in baseline 2 and the more efficient utilization of skewed edges by our 2D-bounded temporal routing. 
\revise{We further demonstrate the importance of the extended IR by enabling and disabling 2D-bounded temporal routing in our framework. Table~\ref{tab:braiding_effect} shows the program depths of 36-qubit benchmarks when 2D size is $6\times6$, with the length limit of temporal edges $D_f$ varying from 5 to 105 (the results of $D_f=1$ can be found in Table.~\ref{tab:secondary_table}). It can be seen that in addition to an average \todoo{$4.08\times$} reduction in 1D depth,
the incorporation of 2D-bounded temporal routing also reduces the required length limit of temporal edges.
Without 2D-bounded temporal routing on the extended IR, the compilation would get stuck at the level of $D_f \sim$ 75 layers due to node congestion. In contrast, with 2D-bounded temporal routing, our compiler can reduce this length limit $D_f$ to 1, as shown in Table~\ref{tab:secondary_table}.}

\begin{table}[h]
    \centering
      \caption{\revise{With and without 2D-bounded temporal routing.}}
      
    \resizebox{\linewidth}{!}{
        \renewcommand*{\arraystretch}{1}
        \begin{normalsize}
            \begin{tabular}{|p{1.8cm}|p{1.5cm}|p{2.0cm}|p{1.9cm}|p{0.9cm}|}
        
        \hline
        Length Limit  $D_f$ of temporal Edges  & Benchmark -\#Qubit & Depth (w/o 2D-bouded temporal routing) & Depth  (w/ 2D-bouded temporal routing)  & Improv. \\
        \hline
        \multicolumn{1}{|c|}{\multirow{3}{*}{\parbox{2cm}{ 5  layers}}}
        & QAOA-36 & - & 245 & - \\
        \cline{2-5}
        & RCA-36 & - & 879 & - \\
        \cline{2-5}
        & VQE-36 & - & 205 & - \\
        \hline
        \multicolumn{1}{|c|}{\multirow{3}{*}{\parbox{2cm}{ 75  layers}}}
        & QAOA-36 & - & 204 & - \\
        \cline{2-5}
        & RCA-36 & 1,902 & 785 & 2.42 \\
        \cline{2-5}
        & VQE-36 & - & 191 & - \\
        \hline
        \multicolumn{1}{|c|}{\multirow{3}{*}{\parbox{2cm}{ 105  layers}}}
        & QAOA-36 & 1,057 & 203 & 5.21 \\
        \cline{2-5}
        & RCA-36 & 1,601 & 743 & 2.15 \\
        \cline{2-5}
        & VQE-36 & 913 & 187 & 4.88 \\
        \hline
        
        \end{tabular}
        \end{normalsize}
    }
    \label{tab:braiding_effect}
\end{table}

\begin{figure*}[tp!]
        \centering
        \includegraphics[width=0.32\linewidth]{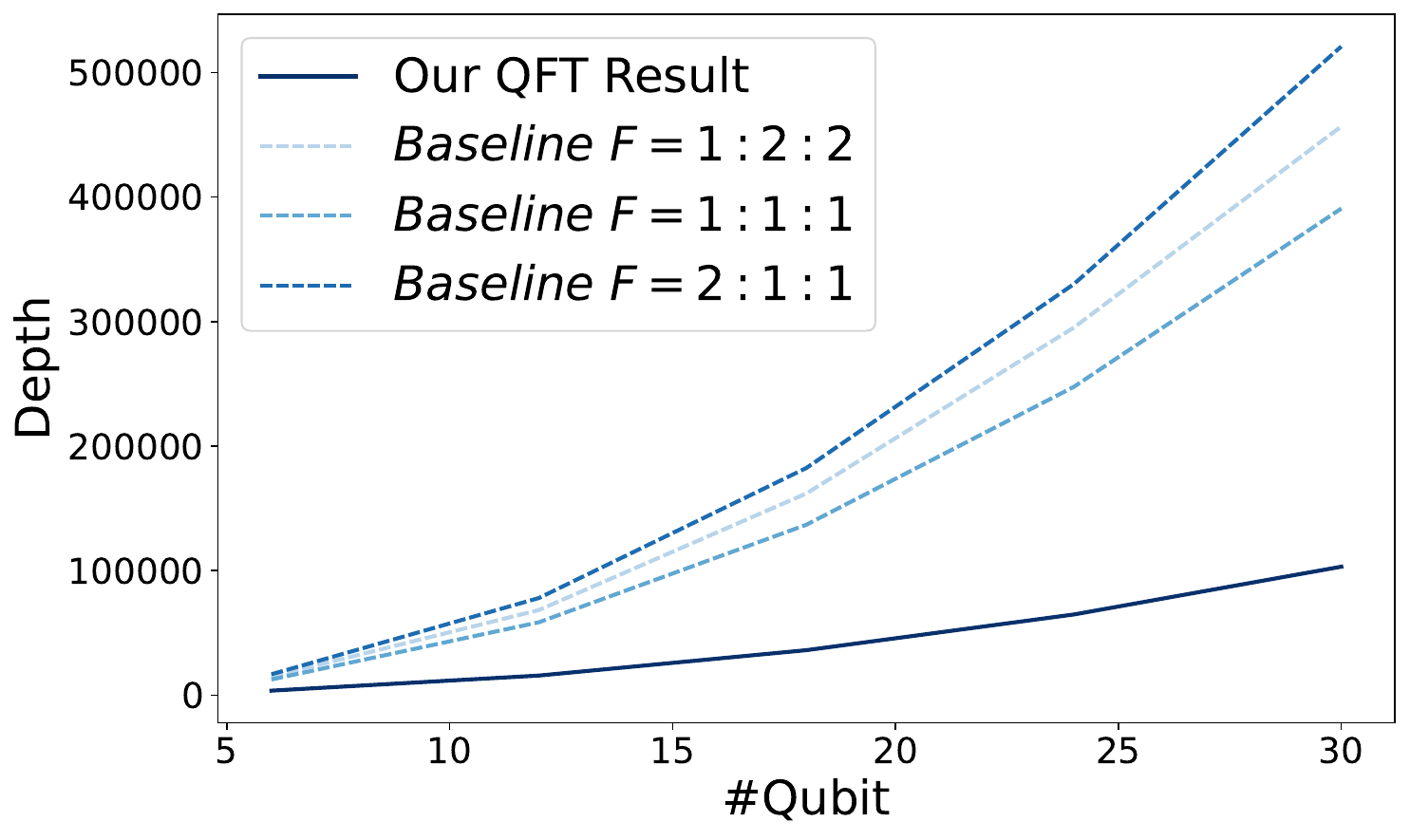}
        \includegraphics[width=0.32\linewidth]{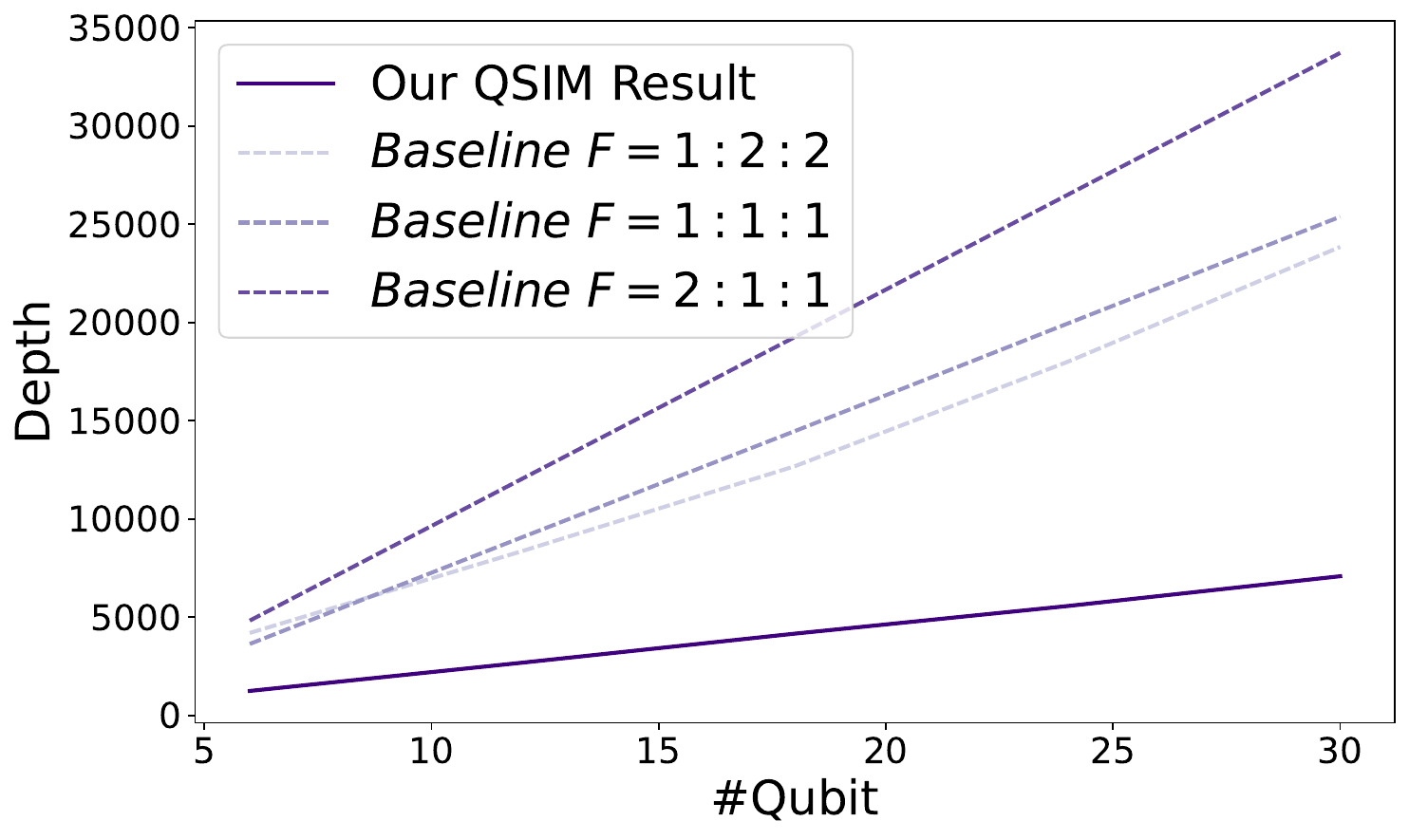}
        \includegraphics[width=0.32\linewidth]{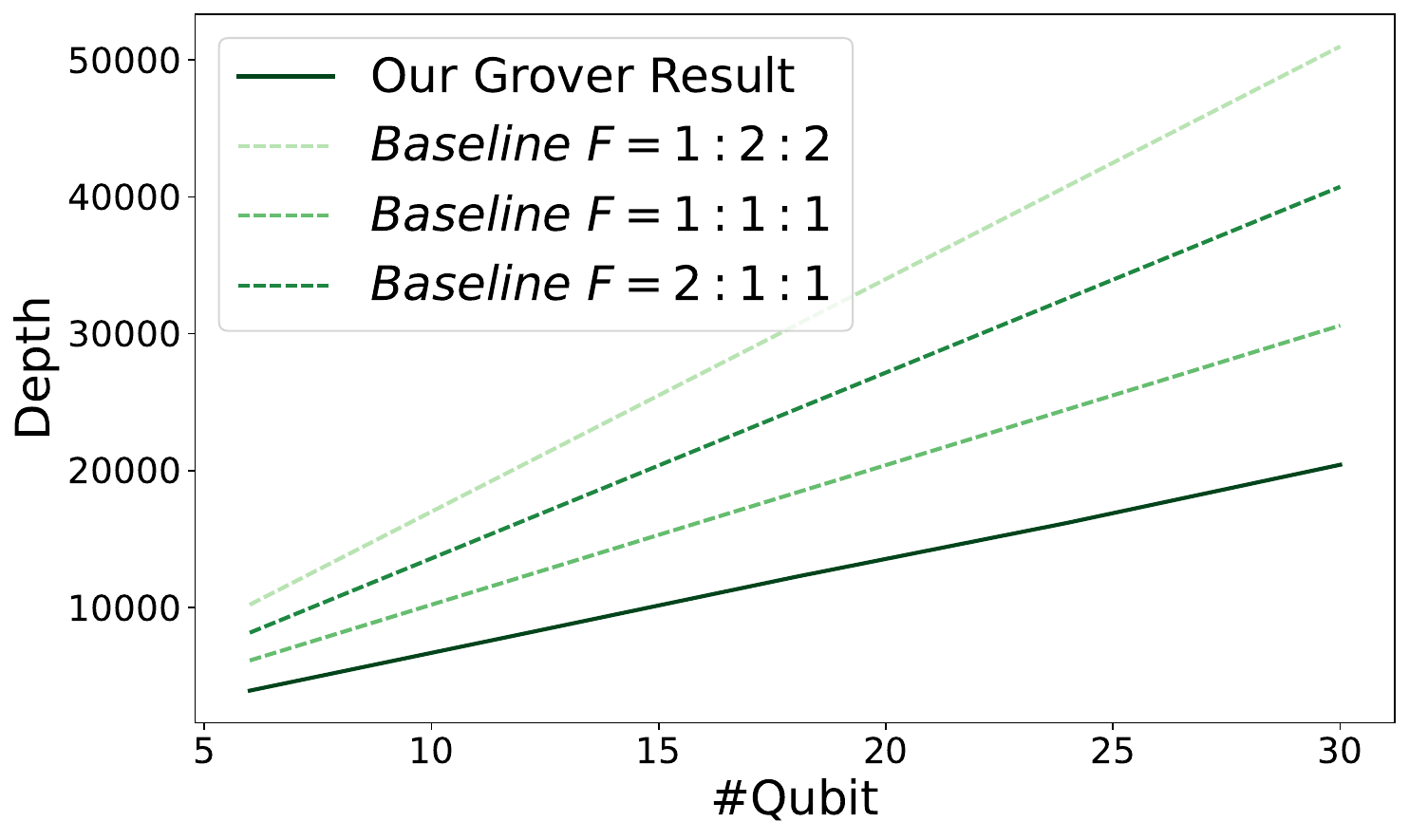}
        \hspace{10pt}(a)\hspace{160pt}(b)\hspace{160pt}(c)
        \caption{Effect of dynamic refresh when QEC is incorporated.}
        \label{fig:qec_eval}
\end{figure*}

\subsection{Incorporation of QEC}
The dynamic refresh in our compiler can also play an important role in FTQC. 
We demonstrate its effect in FTQC by comparing to a static strategy that interleaves QEC patches uniformly over different RSLs, as illustrated in Fig.~\ref{fig:QEC_dynamic}(b).

We adopt the existing scheme of integrating QEC into MBQC on fusion-based architectures \cite{fbqc,FBQC_logical}, using the surface code as it is one of the most promising QEC codes \cite{surface_code}, with a code distance of \todoo{$d=20$} based on previous work \cite{FBQC_logical}. 
We decompose programs into the Clifford + $T$ basis \cite{nielsen2001quantum}, implementing logical operations with lattice surgery \cite{game,high_perform_large_surface}, along with ancilla qubits to facilitate logical CNOT gates \cite{lattice_surgery} and magic state factories to facilitate logical $T$ gates \cite{low_overhead, edge_disjoint}.
Therefore, the logical qubits include algorithmic qubits required by the quantum program, ancilla qubits and magic states.
We vary their ratio from 2:1:1 to 1:1:1 and 1:2:2.

The dynamic refresh mechanism can be easily extended to integrate with QEC, by replacing physical qubits in NISQ with logical blocks in FTQC.
With dynamic refresh, we allocate RSLs for different logical qubits based on their storage time in delay lines and the different logical operations they are involved in. 
Specifically, we divide logical  qubits into active and inactive ones, with the active ones including algorithmic qubits, ancilla qubits and magic states that are currently under logical operations such as lattice surgery and magic state distillation. The dynamic refresh is implemented as the following: (1) Each algorithmic qubit (both active and inactive) and active qubits of other types has to be refreshed within every $D_f$ layers, with $D_f$ depending on the noise level in the delay lines and the number of delay lines available on the hardware. (2) When refresh is not needed by any logical qubits, RSLs are allocated to active logical qubits, with the allocation prioritized for logical qubits with longer storage time in delay lines.

\textbf{Reduction in 1D Depth}
Fig.~\ref{fig:qec_eval} shows the number of consumed RSLs across benchmarks of varying sizes, with and without dynamic refresh. For QFT and Grover, the baseline circuit depth first decreases and then increases as the ratio of different logical qubit types increases, reaching a minimum at the 1:1:1 configuration. For QSIM, the depth is minimized at a 1:2:2 ratio, with the result of 1:1:1 being comparable. Hence we configure the refresh period $D_f$ in our dynamic refresh to match the 1:1:1 setting. It can be seen that even when compared against the best-performing baseline configuration, our technique achieves an average depth reduction of \todoo{2.87}$\times$.


\begin{figure}[h!]
        \centering
        \includegraphics[width=0.7\linewidth]{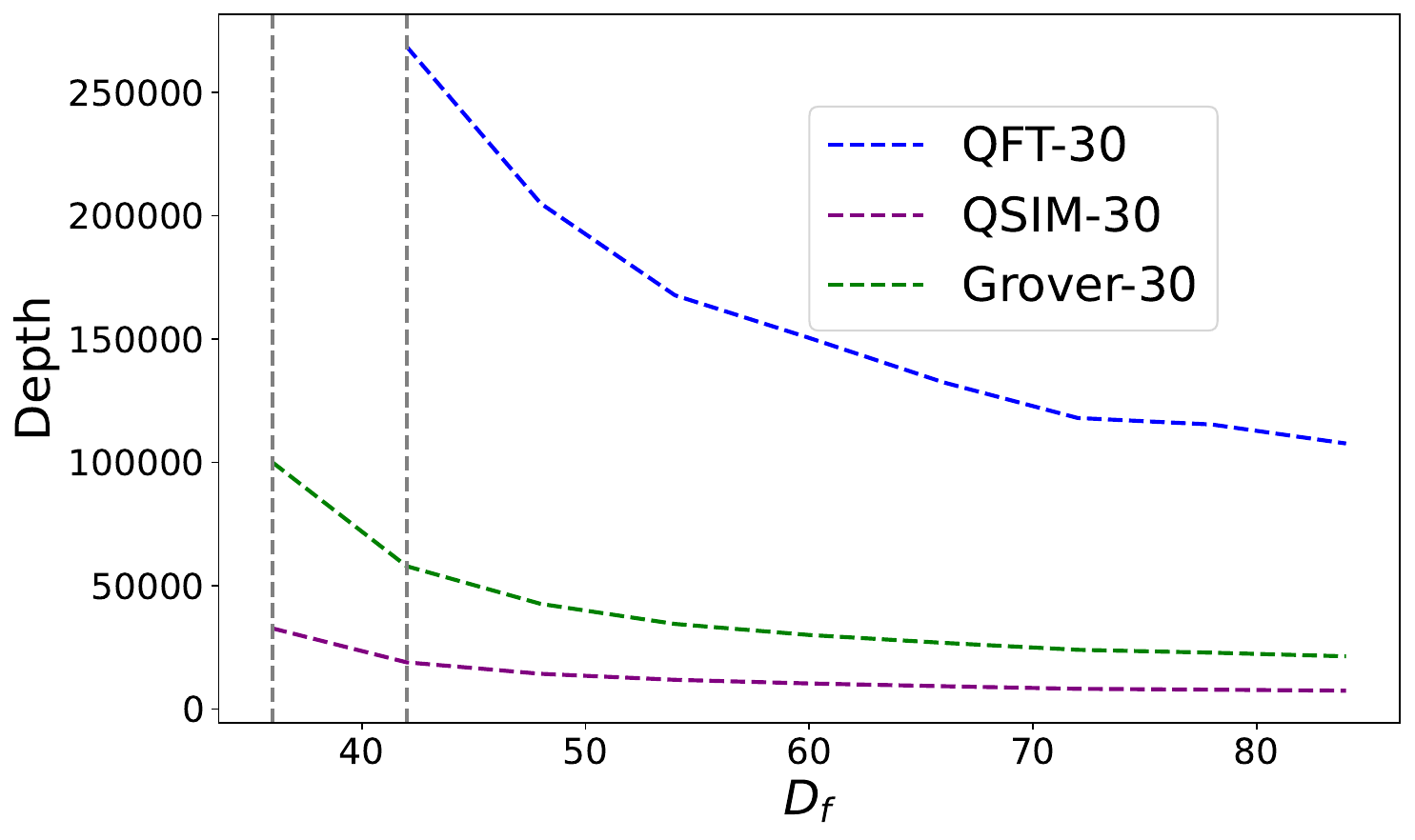}

        \caption{Effects of refresh period limit on QEC depth.}
        \label{fig:qec_refresh_rate}
\end{figure}

\textbf{Effect of Refresh Period}
Fig.~\ref{fig:qec_refresh_rate} demonstrates the adaptability of our dynamic refresh to varying refresh periods. For a 30-qubit circuit, the baseline requires a refresh period $D_f$ of at least 90 to operate. In contrast, it can be seen from Fig.~\ref{fig:qec_refresh_rate} that our dynamic refresh remains effective at significantly lower $D_f$ values around 40, while trading off with a moderate increase in 1D depth as $D_f$ decreases.



\section{Conclusion}

In this work, we present a resource-adaptive compiler for photonic one-way quantum computing. At its core is a novel intermediate representation (IR), which enables two key optimizations: dynamic refresh and 2D-bounded temporal routing. Our evaluation shows that, compared to FlexLattice IR compilers, our compiler reduces the 1D depth by $3.68\times$ while constraining temporal edge lengths within an adaptive bound (as low as 1). Compared to cluster state IR compilers, it can
achieve a $3.56\times$ 1D depth reduction or significantly reduce the 2D size (e.g., from $8 \times 8$ to $3 \times 3$ for 64-qubit programs). Additionally, when applied to FTQC
with surface code, the dynamic refresh reduces the 1D depth by a factor of $2.87\times$.


\begin{acks}

We thank the anonymous reviewers for their constructive feedback. This work is supported in part by NSF 2048144, NSF 2422169, NSF 2427109. 
This material is based upon work supported by the U.S. Department of Energy, Office of Science, National Quantum Information Science Research Centers, Quantum Science Center (QSC). This research used resources of the Oak Ridge Leadership Computing Facility, which is a DOE Office of Science User Facility supported under Contract DE-AC05-00OR22725. 
The Pacific Northwest National Laboratory is operated by Battelle for the U.S. Department of Energy under Contract DE-AC05-76RL01830.
\end{acks}


\bibliographystyle{unsrturl}
\bibliography{references}

\end{document}